\documentclass[aps,prd,preprint,eqsecnum,nofootinbib]{revtex4}
\usepackage{graphicx,epsf}
\newcommand{\be}{\begin{equation}}
\newcommand{\ee}{\end{equation}}

\newcommand{\bea}{\begin{eqnarray}}
\newcommand{\eea}{\end{eqnarray}}
\newcommand{\Tr}{{\rm Tr}}
\newcommand{\nn}{\nonumber}

\def\d{{\rm d}}

\newif\ifdraft
\drafttrue
\newif\ifpreprint
\preprinttrue

\def\sect#1{section~{\ref{#1}}}
\def\sects#1#2{sections~{\ref{#1}} and~{\ref{#2}}}

\def\fig#1{fig.~{\ref{#1}}}
\def\figs#1#2{figs.~{\ref{#1}} and~{\ref{#2}}}
\def\Fig#1{Figure~{\ref{#1}}}

\def\tree{{\rm tree}}

\def\Tr{\, {\rm Tr}}
\def\tr{\, {\rm tr}}

\def\NeqFour{{\cal N}=4}
\def\NeqOne{{\cal N}=1}
\def\NeqEight{{\cal N}=8}

\def\spa#1.#2{\left\langle#1\,#2\right\rangle}
\def\spb#1.#2{\left[#1\,#2\right]}
\def\sand#1.#2.#3{%
\left\langle\smash{#1}{\vphantom1}^{-}\right|{#2}%
\left|\smash{#3}{\vphantom1}^{-}\right\rangle}
\def\sandp#1.#2.#3{%
\left\langle\smash{#1}{\vphantom1}^{-}\right|{#2}%
\left|\smash{#3}{\vphantom1}^{+}\right\rangle}
\def\sandpp#1.#2.#3{%
\left\langle\smash{#1}{\vphantom1}^{+}\right|{#2}%
\left|\smash{#3}{\vphantom1}^{+}\right\rangle}
\def\sandpm#1.#2.#3{%
\left\langle\smash{#1}{\vphantom1}^{+}\right|{#2}%
\left|\smash{#3}{\vphantom1}^{-}\right\rangle}
\def\sandmp#1.#2.#3{%
\left\langle\smash{#1}{\vphantom1}^{-}\right|{#2}%
\left|\smash{#3}{\vphantom1}^{+}\right\rangle}
\def\sandmm#1.#2.#3{%
\left\langle\smash{#1}{\vphantom1}^{-}\right|{#2}%
\left|\smash{#3}{\vphantom1}^{-}\right\rangle}
\def\spab#1.#2.#3{\sandmm#1.#2.#3}
\def\spba#1.#2.#3{\sandpp#1.#2.#3}
\def\spaa#1.#2.#3.#4{\sandmp#1.{#2#3}.#4}
\def\spbb#1.#2.#3.#4{\sandpm#1.{#2#3}.#4}
\def\spash#1.#2{\spa{\smash{#1}}.{\smash{#2}}}

\newbox\charbox
\newbox\slabox
\def\s#1{{      
        \setbox\charbox=\hbox{$#1$}
        \setbox\slabox=\hbox{$/$}
        \dimen\charbox=\ht\slabox
        \advance\dimen\charbox by -\dp\slabox
        \advance\dimen\charbox by -\ht\charbox
        \advance\dimen\charbox by \dp\charbox
        \divide\dimen\charbox by 2
        \raise-\dimen\charbox\hbox to \wd\charbox{\hss/\hss}
        \llap{$#1$}
}}
\def\Circ#1{{
\setbox\charbox=\hbox{$#1$}
\setbox\slabox=\hbox{$\bigcirc$}
        \dimen\charbox=\ht\slabox
        \advance\dimen\charbox by -\dp\slabox
        \advance\dimen\charbox by -\ht\charbox
        \advance\dimen\charbox by \dp\charbox
        \divide\dimen\charbox by 2
\rlap{$\bigcirc$}{\raise-\dimen\charbox\hbox to \wd\slabox{\hss$#1$\hss}}
}}

\def\eqn#1{eq.~(\ref{#1})}

\def\eqns#1#2{eqs.~(\ref{#1}) and~(\ref{#2})}

\def\qb{{\overline {\kern-0.7pt q\kern -0.7pt}}}

\def\e{\epsilon}
\def\ep{\epsilon}
\def\eps{\epsilon}

\def\sign{{\mathop{\rm sign}\nolimits}}

\def\oneloop{{(1)}}
\def\twoloop{{(2)}}
\def\threeloop{{(3)}}
\def\fourloop{{(4)}}
\def\fiveloop{{(5)}}

\def\Ord{{\cal O}}

\def\tlambda{\widetilde{\lambda}}

\def\nn{\nonumber}

\def\MinusVertex{\hbox{\null \hskip -2.8 mm \Circ- 
                      \hskip -2.7mm \null  }}
\def\PlusVertex{\Circ+}
\def\PlusVertex{\hbox{\null \hskip -2.8mm  \Circ+ 
                     \hskip -2.7mm \null }}
\def\tag#1{\tree}

\def\sandp#1.#2.#3{%
\left\langle\smash{#1}{\vphantom1}^{+}\right|{#2}%
\left|\smash{#3}{\vphantom1}^{+}\right\rangle}
\def\ksl{\s{k}}
\def\lsl{\s{l}}

\newbox\ourfigbox
\def\SizedFigureWithCaption#1#2#3{%
\setbox\ourfigbox
  \hbox{\hss\epsfxsize #1 \epsfbox{#2}\hss}
\hbox to \wd\ourfigbox{\vbox{\noindent\copy\ourfigbox\break
\vskip -6mm      \hbox to \wd\ourfigbox{\hss#3\hss}}}
}

\def\spa#1.#2{\left\langle#1\,#2\right\rangle}
\def\spb#1.#2{\left[#1\,#2\right]}
\def\lor#1.#2{\left(#1\,#2\right)}
\def\sand#1.#2.#3{%
\left\langle\smash{#1}{\vphantom1}^{-}\right|{#2}%
\left|\smash{#3}{\vphantom1}^{-}\right\rangle}
\def\sandpp#1.#2.#3{%
\left\langle\smash{#1}{\vphantom1}^{+}\right|{#2}%
\left|\smash{#3}{\vphantom1}^{+}\right\rangle}
\def\sandpm#1.#2.#3{%
\left\langle\smash{#1}{\vphantom1}^{+}\right|{#2}%
\left|\smash{#3}{\vphantom1}^{-}\right\rangle}
\def\sandmp#1.#2.#3{%
\left\langle\smash{#1}{\vphantom1}^{-}\right|{#2}%
\left|\smash{#3}{\vphantom1}^{+}\right\rangle}

\begin{document}
\hfuzz 15 pt


\ifpreprint \noindent
UCLA/07/TEP/04 \hfill ZU-TH 12/07
\hfill Saclay/SPhT--T07/050 \hfill
\fi

\title{Maximally Supersymmetric Planar Yang-Mills Amplitudes
                       at Five Loops}%

\author{Z.~Bern, J.~J.~M.~Carrasco and H.~Johansson}
\affiliation{{} Department of Physics and Astronomy, UCLA\\
\hbox{Los Angeles, CA 90095--1547, USA} }

\author{D.~A.~Kosower}
\affiliation{Service de Physique Th\'eorique,
   CEA--Saclay\\
        F--91191 Gif-sur-Yvette cedex, France\\
{\rm and}\\
Institut f\"ur Theoretische Physik, Universit\"at Z\"urich,\\
        CH--8057  Z\"urich, Switzerland         \\
}

\date{May, 2007}

\begin{abstract}

We present an ansatz for the planar five-loop four-point amplitude in
maximally supersymmetric Yang-Mills theory in terms of loop integrals.
This ansatz exploits the recently observed correspondence between
integrals with simple conformal properties and those found in the
four-point amplitudes of the theory through four loops. We explain how
to identify all such integrals systematically.  We make use of
generalized unitarity in both four and $D$ dimensions to determine the
coefficients of each of these integrals in the amplitude.  Maximal
cuts, in which we cut all propagators of a given integral, are an
especially effective means for determining these coefficients.  The
set of integrals and coefficients determined here will be useful for
computing the five-loop cusp anomalous dimension of the theory which
is of interest for non-trivial checks of the AdS/CFT duality
conjecture.  It will also be useful for checking a conjecture that the
amplitudes have an iterative structure allowing for their all-loop
resummation, whose link to a recent string-side computation by Alday
and Maldacena opens a new venue for quantitative AdS/CFT comparisons.
\end{abstract}

\maketitle

\renewcommand{\thefootnote}{\arabic{footnote}}
\setcounter{footnote}{0}

\section{Introduction}
\label{IntroSection}

Maximally supersymmetric Yang-Mills theory (MSYM) is an important
arena for exploring the properties of gauge theories.  The Maldacena
weak-strong duality~\cite{Maldacena} between MSYM and string theory in
AdS$_5\times S^5$ provides an explicit realization of 't~Hooft's old
dream~\cite{tHooft} of expressing the strongly coupled limit of a
gauge theory in terms of a string theory.  In addition, the
higher-loop planar space-time scattering amplitudes of MSYM appear to
have a remarkably simple and novel iterative
structure~\cite{ABDK,BDS}.  This structure allows higher-loop
amplitudes to be expressed in terms of lower-loop amplitudes.  This
simplicity may well be connected to the observed integrability of the
theory in the planar limit~\cite{Integrability,EdenStaudacher,BES}.

The iterative structure of the planar amplitudes was first proposed in
ref.~\cite{ABDK}, and confirmed for the two-loop four-point amplitude.
An independent two-loop check was given in ref.~\cite{HiddenBeauty}.
In ref.~\cite{BDS}, this proposal was fleshed out for all planar
maximally helicity violating (MHV) amplitudes, expressed via a
specific all-loop exponentiation formula which was confirmed for
three-loop four-point amplitudes.  The iteration formula has also been
shown to hold for two-loop five-point amplitudes~\cite{FivePtTwoLoop}.
At four loops the amplitudes are known in terms of a set of integrals,
but the integrals themselves have not yet been evaluated
fully~\cite{BCDKS,CSVFourLoop}.  In this paper, we provide the
corresponding integral representation of the five-loop four-point
planar amplitude, for use in future studies of its properties.  In a
very recent paper, Alday and Maldacena~\cite{AldayMaldacena} have
shown how to perform a string-side computation of the same gluon
amplitudes in the strong-coupling limit.  This opens new and exciting
possibilities of quantitative checks of the AdS/CFT correspondence,
going beyond anomalous dimensions to detailed dependence on
kinematics.

In addition to exhibiting an iterative structure, the scattering
amplitudes provide new and nontrivial information on the AdS/CFT
correspondence.  Using considerations of integrability, an integral
equation for the cusp (soft) anomalous dimension --- valid to all loop
orders --- was written down by Eden and
Staudacher~\cite{EdenStaudacher}.  This equation agreed with the first
three loop orders~\cite{KLOV,BDS}, but its reliance on various
assumptions cast doubt on whether it would hold beyond this.  We now
know that this original proposal requires modification because of the
recent calculation of the four-loop cusp anomalous dimension from the
infrared singular terms of a four-loop MSYM
amplitude~\cite{BCDKS,CSVFourLoop}. A remarkable new
integral equation proposed by Beisert, Eden and Staudacher
(BES)~\cite{BES} is in agreement with this calculation.
Surprisingly, the first four loop orders of the planar cusp anomalous
dimension contain sufficient information to test the AdS/CFT
correspondence to the level of a few percent~\cite{BCDKS}, using the
interpolating function technique introduced by Kotikov, Lipatov, and
Velizhanin~\cite{KLV} as well as Pad\'e approximants. This provides an
independent guess of the entire perturbative series~\cite{BCDKS},
matching the one generated by the BES integral equation.  Detailed
studies of the BES equation~\cite{Klebanov} confirm that it has the
proper behavior~\cite{StrongCouplingLeading} at strong coupling,
giving a high degree of confidence in it.  Nonetheless, further checks
on the perturbative side would be quite valuable.  The first such
check requires a computation of the five-loop anomalous dimension,
which --- following the approach taken at four loops --- requires an
expression for the five-loop four-point amplitude.

The unitarity method~\cite{NeqFourOneLoop,Fusing, DDimUnitarity,
OneLoopReview,DimShift,GeneralizedUnitarity} provides a powerful
method for computing gauge and gravity loop amplitudes and has played
a central role in obtaining MSYM loop
amplitudes~\cite{NeqFourOneLoop, Fusing, BRY, BDDPR, ABDK, BDS, BRYProceedings,
FivePtTwoLoop,BCDKS}.
An important recent improvement~\cite{BCFUnitarity} is the use of
complex momenta~\cite{WittenTopologicalString} within the framework of
generalized unitarity~\cite{GeneralizedUnitarity}. In particular, this
allows one to define a non-zero massless three-point amplitude,
which vanishes for real momenta.  At one loop this enables an easy
algebraic determination of the coefficient of any box integral
appearing in the theory, because the cut conditions freeze the loop
integrals~\cite{BCFUnitarity}.  Some of these ideas have also been
applied at two loops~\cite{BuchbinderCachazo}.  In this paper we apply
these ideas to develop a {\it maximal-cutting} method for efficiently
determining coefficients of higher-loop integrals.

In the MSYM theory, a special set of cuts --- the iterated two-particle
cuts --- give rise to the ``rung rule'' for systematically obtaining
higher-loop integral representations of planar four-point
amplitudes~\cite{BRY,BDDPR}.  At two and three loops this rule
generates all contributions appearing in the amplitudes.  However,
starting at four loops, new integrals arise which are not generated
by the rung rule.  In ref.~\cite{BCDKS}, these were computed
explicitly using generalized unitarity by relying on a set of mild
assumptions.

These integrals, along with two others that do not appear in the
amplitude, are predicted by a procedure relying on a beautiful
observation due to Drummond, Henn, Smirnov and Sokatchev
(DHSS)~\cite{DHSS}.  These authors noticed that, through three loops,
the massless integrals appearing in the planar four-point amplitude
are in direct correspondence with conformally invariant integrals. This
correspondence comes from replacing dimensional regularization with an
off-shell infrared regularization in four dimensions.  We shall call
the dimensional regularized version of the conformal integrals {\it
pseudo-conformal}, since dimensional regularization breaks
conformal invariance.  DHSS also gave simple rules for generating all
such integrals via ``dual diagrams''.  The direct evaluation of
generalized unitarity cuts confirmed~\cite{BCDKS} at four loops, that
only such pseudo-conformal integrals appear in the planar amplitude.
In the present work, we will assume that this is also true at higher
loops, beginning with five loops.  This provides a basis set of
integrals for the planar (leading-color) contributions to the
five-loop amplitude.  We then use the unitarity method to determine
the coefficients of these integrals in the planar five-loop four-point
amplitude as well as to provide consistency checks on the absence of
other integrals.

We make use of an additional observation: the cutting equations hold
at the level of the integrands, prior to carrying out any loop
integrals.
Indeed they hold independently for each of the multiple 
solutions of the cutting equations. 
These properties are
especially powerful when combined with the basis of pseudo-conformal
integrals. The problem is then reduced to an algebraic problem of
determining the coefficient of each integral.  Remarkably, it turns
out that after dividing by the tree amplitude, the coefficients are
pure numbers taking on the values $-1,0,$ or~$1$. This property is
already known to hold for the four-point amplitude through four
loops~\cite{BCDKS}, and here we confirm it through five loops.

At one loop, complete dimensionally regularized amplitudes in the MSYM
theory can be constructed using only four-dimensional helicity
amplitudes, greatly simplifying their
construction~\cite{NeqFourOneLoop,Fusing}.  Unfortunately, no such
theorem exists at higher loops.  Any rigorous construction of
amplitudes requires that $D$-dimensional momenta be used in the cuts.
It is worth noting that if our assumption of a pseudo-conformal basis
of integrals is correct, with dimension independent coefficients, then
unitarity in four dimensions {\it does} suffice to determine these
amplitudes in all dimensions. Our partial checks of $D$-dimensional
cuts provide non-trivial evidence that this assumption is correct.
This is rather remarkable because away from four dimensions there is
{\it a priori\/} no reason why a simple analytic continuation of the
dimension of the integrals should give correct results.

Our expression for the five-loop four-point planar MSYM
amplitude in terms of integral functions should be useful in a number of
studies.  The infrared singularities present in the amplitude encode
the so-called cusp or soft anomalous dimension. At five loops these
singularities begin at $1/\eps^{10}$ (where $\eps$, as usual, is the
dimensional regularization parameter: $\eps = (4-D)/2$).  Evaluation
of the infrared singular terms through $1/\eps^2$ would allow the
extraction of the five-loop cusp anomalous dimension, as has
been done at three and four loops~\cite{BDS,BCDKS}. An evaluation
though order $1/\eps$ would allow the
extraction of a second anomalous dimension connected to a form factor.
An evaluation through $\Ord(\eps^0)$ would allow a five-loop check of
the iterative structure of the amplitudes, providing strong evidence
that it continues to all loop orders.  Although evaluating loop integrals
is rather challenging, there has been rapid progress using Mellin-Barnes
representations~\cite{VolodyaIntegrals} and in automating the required
analytic continuations~\cite{MB}, allowing for explicit computations
through four loops.  Recently, there has also been
progress in isolating the subsets of terms which determine the anomalous
dimension~\cite{CSVFourLoop}, greatly simplifying its calculation.
Further development will presumably be needed to 
apply these advances to the five-loop amplitude.

Another important reason for studying MSYM amplitudes is their
intimate connection to $\NeqEight$ supergravity amplitudes.  The
identification of additional cancellations in this
theory~\cite{GravityCancel} suggests that in four dimensions it may be
ultraviolet finite to all loop
orders~\cite{GravityFinite,ThreeLoopNEqEight}.  (See also
refs.~\cite{KITPTalk}.)  String dualities also hint at UV finiteness
for $\NeqEight$ supergravity~\cite{StringFinite}, although this is
weakened by issues with towers of light non-perturbative states from
branes wrapped on the compact dimensions~\cite{OoguriPrivate}.
Remarkably, computational advances for gauge theory amplitudes can be
imported~\cite{BDDPR} directly into calculations of gravity
amplitudes, by combining the unitarity method with the
Kawai-Lewellen-Tye~\cite{KLT} tree-level relations between gauge and
gravity theories.  This allows cuts of gravity loop amplitudes to be
expressed as double copies of cuts of corresponding gauge theory
amplitudes~\cite{BDDPR}.  This strategy has recently been applied to
obtain the three-loop four-point amplitude of $\NeqEight$
supergravity, starting from corresponding $\NeqFour$ MSYM
amplitudes~\cite{ThreeLoopNEqEight}.  That computation shows that at
least through three loops, MSYM and $\NeqEight$ supergravity share the
same ultraviolet power-counting.  Because they share the same critical
dimension for ultraviolet finiteness, both are ultraviolet finite in
four dimensions. The five-loop planar super-Yang-Mill amplitudes
obtained here will be an important input for obtaining the
corresponding five-loop supergravity amplitudes.  (The supergravity
calculation also requires the non-planar contributions.)

The paper is organized as follows. In \sect{NotationSection}, we
briefly summarize known properties of the amplitudes and define
the notation used in the remainder of the paper.  In
\sect{ConformalIntegralsSection}, we review the observations of
refs.~\cite{DHSS,BCDKS}, on the exclusive appearance of
pseudo-conformal integrals in
planar four-point amplitudes.  Because candidate integrals
proliferate as the number of loops increases,
we give a systematic procedure in \sect{ConstructionSection} for
constructing these integrals. The results of this
procedure at five loops are given in \sect{ResultSection}, along with
the coefficients of the integrals determined via unitarity.  Our
ansatz for the amplitude is also presented in this section.  We then
briefly review generalized unitarity in \sect{UnitaritySection}. A
description of the cuts used to determine the integral coefficients is
given in sections~\ref{MaximalCutSection} and
\ref{ConfirmingCutsSection}, along with a description of the method of
maximal cuts introduced in this paper. Our conclusions and some
comments on the outlook are given in \sect{ConclusionSection}.

\section{Notation and Review of MSYM Amplitudes}
\label{NotationSection}

We use a standard color decomposition~\cite{TreeReview,OneLoopReview} for
the MSYM amplitudes in order to
disentangle color from kinematics.  In this paper we focus
on the leading-color planar contributions, which have a color structure
similar to those of tree amplitudes, up to overall factors
of the number of colors, $N_c$.  The color-decomposed form of planar
contributions to the $L$-loop $SU(N_c)$ gauge-theory $n$-point
amplitudes is,
\begin{eqnarray}
{\cal A}_n^{(L)} & = & g^{n-2}
 \Biggl[ { 2 e^{- \e \gamma} g^2 N_c \over (4\pi)^{2-\e} } \Biggr]^{L}
 \sum_{\rho}
\Tr( T^{a_{\rho(1)}} T^{a_{\rho(2)}}
   \ldots T^{a_{\rho(n)}} )
               A_n^{(L)}(\rho(1), \rho(2), \ldots, \rho(n))\,, \hskip 2 cm
\label{LeadingColorDecomposition}
\end{eqnarray}
where $A_n^{(L)}$ is an $L$-loop color-ordered partial amplitude.  We
have followed the normalization conventions of ref.~\cite{BDS}.  Here
$\gamma$ is Euler's constant, and the sum runs over non-cyclic
permutations, $\rho$, of the external legs.  In this expression we
have suppressed labels of momenta and helicities, leaving only the
indices identifying the external legs.  Our convention is that all
legs are outgoing. This decomposition holds for all particles in the
gauge super-multiplet.

We also define a loop amplitude normalized by the tree amplitude,
\begin{equation}
M_n^{(L)} \equiv  A_n^{(L)}/A_n^{\tree}\,.
\label{LoopOverTree}
\end{equation}
Supersymmetry Ward identities~\cite{SWI} guarantee that, after
dividing out by the tree amplitudes, MHV amplitudes are identical for
any helicity configuration~\cite{DimShift}. (The complete set of these
tree amplitudes are tabulated in Appendix E of ref.~\cite{BDDPR}.)
Because four-point amplitudes are always maximally helicity violating,
this holds for all four-point amplitudes.  Because it is independent
of the position of the two negative helicity legs, $M_n^{(L)}$ has
complete cyclic and reflection symmetry.  A practical consequence of
this is that once a coefficient of a given integral is determined, the
coefficient of integrals related by cyclic or reflection symmetry
follow trivially.

In our evaluations of four-dimensional unitarity cuts, we use the
spinor helicity formalism~\cite{SpinorHelicity,TreeReview}, in which
the amplitudes are expressed in terms of spinor inner products,
\begin{eqnarray}
&& \spa{j}.{l} = \langle j^- | l^+ \rangle = {\overline u_-(k_j)} u_+(k_l)\,,
\hskip 2 cm
\spb{j}.{l} = \langle j^+ | l^- \rangle = {\overline u_+(k_j)} u_-(k_l)\,,
\nn \\
&& \sand{a}.{k_b + k_c}.{d} = \overline u_-(k_a) [\ksl_b +
\ksl_c]  u_-(k_d)\ \label{spinorproddef}
\end{eqnarray}
where
$u_\pm(k)$ is a massless Weyl spinor with momentum $k$ and positive
or negative chirality. Our conventions follow the QCD literature,
with $\spb{i}.{j} = \sign(k_i^0 k_j^0)\spa{j}.{i}^*$ for real
momenta so that,
\begin{equation}
\spa{i}.{j} \spb{j}.{i} = 2 k_i \cdot k_j = s_{ij}\,.
\end{equation}
We also define,
\begin{equation}
 \lambda_{k_i} \equiv u_+(k_i), \qquad \tlambda_{k_i} \equiv u_-(k_i) \,.
\label{lambdadef}
\end{equation}
For complex momenta these two spinors are independent, though
they are dependent for real momenta.

In ref.~\cite{ABDK}, a conjecture was presented that MSYM amplitudes
possess an iterative structure, based on an observed iteration of
two-loop four-point amplitudes.
In ref.~\cite{BDS}, this was fleshed out for MHV amplitudes
into an explicit exponentiation ansatz to all loop orders.
 Through five loops, the expansion of the exponential gives
the iteration relations,
\begin{eqnarray}
M_n^{\twoloop}(\e)
&=& {1 \over 2} \Bigl(M_n^{\oneloop}(\e) \Bigr)^2
 + f^\twoloop(\e) \, M_n^{\oneloop}(2\e) + C^{(2)}
 + \Ord(\e)\,,
\label{TwoLoopIteration} \\
M_n^\threeloop(\e) &=& - {1\over 3} \Bigl[M_n^\oneloop(\e)\Bigr]^3
            + M_n^\oneloop(\e)\, M_n^\twoloop(\e)
            + f^\threeloop(\e) \, M_n^\oneloop (3\,\e) + C^{(3)}
            + \Ord(\e) \,,
\label{ThreeLoopteration} \\
M_n^\fourloop(\e) &=&  {1\over 4} \Bigl[M_n^\oneloop(\e)\Bigr]^4
            - \Bigl[M_n^\oneloop(\e)\Bigr]^2  M_n^\twoloop(\e)
            + M_n^\oneloop(\e)  M_n^\threeloop(\e)
            + {1\over 2} \Bigl[M_n^\twoloop(\e)\Bigr]^2 \nn\\
&& \hskip 3 cm \null
            + f^\fourloop(\e) \, M_n^\oneloop (4\,\e) + C^{(4)}
            + \Ord(\e) \,,
\label{FourLoopIteration} \\
M_n^\fiveloop(\e) &=&  -{1\over 5} \Bigl[M_n^\oneloop(\e)\Bigr]^5
            + \Bigl[M_n^\oneloop(\e)\Bigr]^3  M_n^\twoloop(\e)
            - \Bigl[M_n^\oneloop(\e)\Bigr]^2  M_n^\threeloop(\e)
            - M_n^\oneloop(\e) \Bigl[M_n^\twoloop(\e) \Bigr]^2  \nn\\
&&  \null
            + M_n^\oneloop(\e) M_n^\fourloop(\e)
            + M_n^\twoloop(\e) M_n^\threeloop(\e)
            + f^\fiveloop(\e) \, M_n^\oneloop (5\,\e) + C^{(5)}
            + \Ord(\e) \,,
\label{FiveLoopIteration}
\end{eqnarray}
where $f^{(L)}(\e)$ is a three term series in $\eps$,
\begin{equation}
f^{(L)}(\e) = f_0^{(L)} + \e f_1^{(L)} + \e^2 f_2^{(L)} \,,
\end{equation}
and $f_i^{(L)}$ are numbers independent of the kinematics and
of the number of external legs $n$.  Similarly, the $C^{(L)}$ are
also pure numbers.  The constant  $f_0^{(L)}$ is proportional
to the $L$-loop
cusp (soft) anomalous dimension $\gamma_K^{(L)}$,
\begin{equation}
f_0^{(L)} = {1\over 4 } \gamma_K^{(L)}\,.
\end{equation}
After subtracting the known infrared singularities~\cite{MagneaSterman}
the iteration relation takes on a rather simple exponential form,
\begin{equation}
{\cal F}_n (0) = \exp\biggl[ {1\over 4} \gamma_K F^\oneloop_n (0) + C \biggr]
\end{equation}
where $F_n^\oneloop(0)$ is the $n$-point one-loop finite remainder,
$\gamma_K$ is the complete cusp anomalous dimension, and $C$ depends
on the coupling but not on the external momenta.  Very recently Alday 
and Maldacena have matched this expression at strong coupling for $n=4$
using string theory~\cite{AldayMaldacena}.

The iteration conjecture has so far been confirmed for two- and
three-loop four-point amplitudes~\cite{ABDK,BDS} as well for two-loop
five-point amplitudes~\cite{FivePtTwoLoop}.  For the four-loop
four-point amplitude, the integrand is known~\cite{BCDKS} and has been
shown to generate the correct form of the infrared singularities
though $\Ord(1/\ep^2)$.  This has been used to extract the
four-loop contribution to the cusp anomalous
dimension~\cite{BCDKS,CSVFourLoop} numerically.

As the amplitudes are infrared divergent, we need to regulate them.
In order to preserve the supersymmetry we use the four-dimensional
helicity (FDH) scheme~\cite{FDH}, which is a relative of
Siegel's dimensional reduction scheme~\cite{Siegel}.

\section{Pseudo-Conformal Integrals}
\label{ConformalIntegralsSection}

Conformal properties offer a
simple way to identify integrals that can appear in planar MSYM
amplitudes~\cite{DHSS,BCDKS}.
This observation allows us to easily
identify a basis of integrals, whose coefficients can be
determined via the unitarity method.  This greatly simplifies the cut
analysis because we need to determine only these coefficients
to obtain the planar amplitudes.

Although the underlying theory is conformally invariant, there is as
yet no proof that only integrals dictated by conformal invariance can appear.
One obvious complication to providing such a proof is the infrared
divergence of the amplitudes, and the subsequent need to regulate the
integrals (via dimensional regularization), which breaks the conformal
invariance.  As mentioned in the introduction, we therefore call the
integrals corresponding to conformally-invariant ones 
``pseudo-conformal''.  We shall describe in this
section how to implement this correspondence.
In principle, it is possible
that individual integrals appearing in the amplitude would have no
special conformal properties, yet the complete amplitude would retain
simple conformal properties because of cancellations between
integrals. 
Through four loops~\cite{BCDKS}, however, only pseudo-conformal integrals
appear in the four-point amplitude.  This provides compelling evidence
that this is a general property of MSYM four-point
amplitudes.\footnote{For higher-point amplitudes, the situation is
more complicated, as can be confirmed by checking the conformal
properties of known results~\cite{FivePtTwoLoop} for the five-point
amplitudes at two loops.}

We therefore assume that only pseudo-conformal integrals appear
in the five-loop planar amplitudes.  One way of proving the
correctness of this assumption would be to compute a sufficient number
of cuts in $D$ dimensions to determine the amplitude completely. In
this paper, we discuss only a partial confirmation.

Dimensional
regulation of the infrared singularities breaks the
conformal symmetry.  For the purposes of exposing
the conformal symmetry, we instead regulate the infrared
divergences by taking the external momenta $k_i$ off shell and
letting the dimension be $D=4$.  We will obtain a pseudo-conformal
integral from a suitable conformal integral by reversing this change
of regulator.

The authors of ref.~\cite{DHSS} provide a simple way of
making manifest the conformal properties of planar integrals via
``dual diagrams''.  The dual diagrams provide a direct method of
identifying all
conformally invariant loop integrals.\footnote{The dual diagrams are related
to the dual graphs used in graph theory~\cite{Nakanishi}.}  In
general, conformal properties are not obvious in the momentum-space
representation of loop integrals, but with a simple change
of variables encoded by the dual diagrams we can make these
properties manifest.

\begin{figure}
\centerline{\epsfxsize 4.7 truein \epsfbox{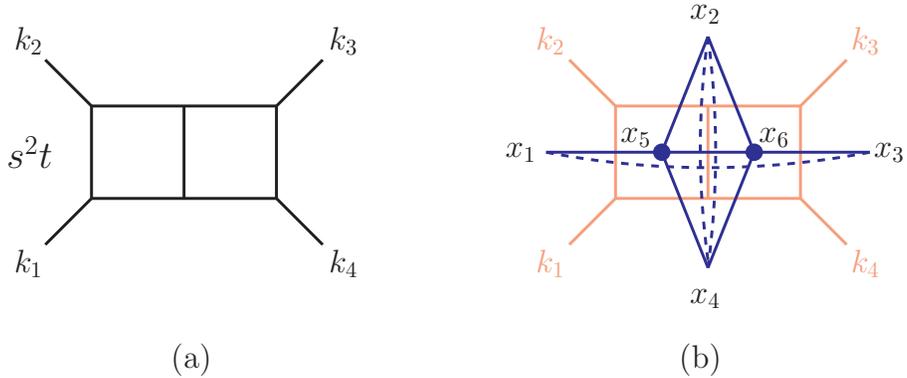}}
\caption{The two-loop planar double-box integral (a) and its dual
(b) overlaying a faded version of (a). In (b) the dashed
lines represent a numerator factor of $(x_{24}^2)^2 x_{13}^2 = s^2
t$. This inserted numerator factor is needed for conformal
invariance of the integral.} \label{dual2loop}
\end{figure}

Let us give an illustrative example.
Consider the two-loop double-box integral of \fig{dual2loop}(a),
\begin{equation}
I^{(2)}(s,t) = (-i e^{\eps\gamma} \pi^{-D/2})^2 s^2 t \int
\frac{{d}^D p \,{d}^D q}{p^2 (p-k_1)^2 (p-k_1-k_2)^2
q^2 (q-k_4)^2 (q-k_3-k_4)^2 (p+q)^2} \,, 
\label{2box}
\end{equation}
where $s=(k_1+k_2)^2$ and $t=(k_2+k_3)^2$.  

After replacing the regulator
as mentioned above,
the conformal symmetry can then be exposed via the change of variables,
\begin{equation}
k_1=x_{41}\,, \hskip 1 cm k_2=x_{12},\hskip 1 cm k_3=x_{23},
\hskip 1 cm k_4=x_{34},\hskip 1 cm p=x_{45}\,, \hskip 1 cm
q=x_{64} \,,
\label{map}
\end{equation}
where $x_{ij} \equiv x_i-x_j$.  The new variables automatically
satisfy momentum conservation
\begin{equation}
x_{41}+x_{12}+x_{23}+x_{34}=0 \hskip .3 cm
\Longleftrightarrow \hskip .3 cm
k_1 + k_2 + k_3 + k_4 = 0 \,.
\end{equation}
After substituting the new variables into~\eqn{2box} and taking
$D=4$, the double-box integral takes on a very symmetric form,
\begin{equation}
I^\twoloop(s, t) = (-i \pi^{-2})^2 x^4_{24} x^2_{13} \int
\frac{{d}^4 x_5\, {d}^4 x_6}{x^2_{45} x^2_{15}
x^2_{25} x^2_{46} x^2_{36} x^2_{62} x^2_{56}} \,.
\label{2boxdual}
\end{equation}
The conformal-invariance properties follow
from examining its behavior under inversion, 
$x^\mu \rightarrow x^\mu/x^2$,
\begin{equation}
x^2_{ij} \rightarrow \frac{x^2_{ij}}{x^2_i x^2_j}\,,  \hskip 2.5 cm
{d}^4x_i \rightarrow \frac{{d}^4x_i }{x^8_i} \,.
\label{ConformalWeight}
\end{equation}
Under this inversion the double-box~(\ref{2boxdual}) is invariant
because each external point $x_1,x_2,x_3,x_4$ appears equally many
times in the numerator as in the denominator, while the internal
points $x_5,x_6$ appears exactly four times in the denominator,
precisely canceling the behavior of the integration measure. The $x$
variables are useful because inversion respects momentum conservation,
which is not true for an inversion of the original momentum variables.

More generally, following refs.~\cite{DHSS,BCDKS}, we keep track of
conformal weights using dual diagrams.  To obtain the dual
representation, start with the momentum representation shown in
\fig{dual2loop}(a) and place internal points $x_5,x_6$ inside each
loop, as well as external points $x_1,x_2,x_3, x_4$ between each
pair of external momenta, as shown in \fig{dual2loop}(b).
Following ref.~\cite{DHSS} we mark the internal integration points by
solid dots at the center of each loop but in most cases leave the
external points unmarked.  Solid lines represent an inverse power of
$x_{ij}^2$, corresponding to the dual propagator $1/x^2_{ij}$, which crosses
exactly one Feynman propagator whose momentum is equal to $x_{ij}$.
Dashed lines represent a positive power of an $x_{ij}^2$, such as
the numerator factors of $s = x_{24}^2$ and $t = x_{13}^2$.  The
$x_{ij}^2$ represented by a dashed line correspond the sum of the
momenta of the ordinary propagator lines they cross. (The dashed
lines can be deformed to cross different propagators, but momentum
conservation ensures that this does not affect the value of the dual
invariants $x^2_{ij}$.)  The dual diagram constructed in the way in
\fig{dual2loop}(b), is in direct correspondence to the dual integral
(\ref{2boxdual}).

We can further restrict the possible set of conformal integrals
by requiring that they have only logarithmic behavior
in the on-shell limit.  That is, we are not interested in conformal
integrals which vanish or diverge with a power-law behavior in any
$k_i^2$, because these do not correspond to massless integrals in
dimensional regularization.  For example, numerator factors
such as $x_{12}^2 = k_2^2$ are not
allowed.  Similarly, factors such as $1/x_{12}^2 = 1/k_2^2$ are
excluded because their power singularities are too
severe for the required logarithmic behavior of infrared
singularities.  We then obtain a pseudo-conformal integral, as
discussed earlier, by replacing the off-shell regulator with
the usual dimensional one.

It is straightforward to generalize this graphical mapping to
any loop order, allowing for a relatively simple bookkeeping
of the change of variables between the momenta and the dual
$x_i$ variables. The map in \eqn{map} exemplifies the convention we use
for external momenta for all diagrams in this paper.

The conformal weights are easy to read off directly from the dual
diagrams. For a dual diagram to be conformally invariant it must satisfy
the following: The number of solid lines minus the number of dashed
lines entering a point $x_i$ must be zero for external points, and four
for internal points. The conformal weight of four for
internal points cancels the conformal weight of the integration
measure given in \eqn{ConformalWeight}.  As observed in
ref.~\cite{BCDKS}, a consequence of requiring integrals to be
conformal is that integrals with triangle or bubble subdiagrams are
not allowed.

\begin{figure}[t]
\centerline{\epsfxsize 4.5 truein \epsfbox{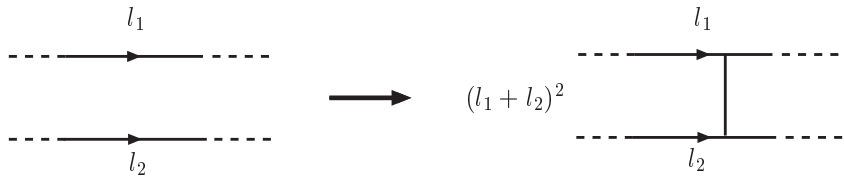}}
\caption{The rung-rule for generating higher-loop
integrands from lower-loop ones.}
\label{RungRuleFigure}
\end{figure}

One class of pseudo-conformal integrals may be understood in terms of the
``rung-rule'' of ref.~\cite{BRY}.  This rule instructs its user to
generate contributions to an $(L+1)$-loop amplitude from a known
$L$-loop amplitude by inserting a new leg between each possible pair
of internal legs, as shown in \fig{RungRuleFigure}. From this set,
all diagrams with either triangle or bubble subdiagrams are removed.
The new loop momentum is integrated over, after including an
additional factor of $(l_1 + l_2)^2$ in the numerator, where $l_1$
and $l_2$ are the momenta flowing through the indicated lines. (With
the conventions used here it is convenient to remove a factor of $i$
from the numerator factor, compared to ref.~\cite{BRY}.) Each
distinct contribution should be counted only once, even if it
can be generated in multiple ways.  Contributions
arising from identical diagrams (that is, having identical propagators)
but with
distinct numerators count as distinct contributions.  The
diagrams obtained by iterating this procedure are sometimes called Mondrian
diagrams, because of their similarity to Mondrian's art.

\begin{figure}[t]
\centerline{\epsfxsize 4.5 truein \epsfbox{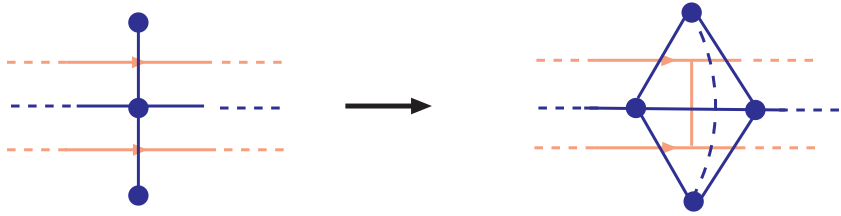}}
\caption{The rung rule maintains conformal weight.  If the dual diagram
prior to applying the rung rule has the proper conformal weight
so will the resulting diagram.}
\label{RungRuleDualFigure}
\end{figure}

The rung rule may be understood as a consequence of the conformal
properties of the integrals.  As illustrated in
\fig{RungRuleDualFigure}, if the starting integral is
pseudo-conformal, inserting a rung splits the inner loop into two
side-by-side loops, and the conformal weight of the central dots in
the figure is unchanged.  However, the upper and lower loop need an
additional dashed line connecting their central dots to maintain their
conformal weight.  This dashed line corresponds exactly to the
factor of $(l_1+l_2)^2$ required by the rung rule in
\fig{RungRuleFigure}.

The rung-rule, unfortunately, does not generate the complete set of
planar integrals~\cite{BCDKS}.  However, at least through four loops
any diagram that it generates is obtained with the correct sign.  Here
we confirm this observation at five loops, using the unitarity
method. To obtain the remaining, non-rung-rule contributions to the
five-loop amplitude, we start with the other pseudo-conformal
integrals and determine their coefficients using the unitarity method.

\section{Generating the Planar Pseudo-Conformal Diagrams}
\label{ConstructionSection}

The proliferation of candidate pseudo-conformal integrals with
increasing number of loops encourages the development of a systematic
construction procedure. One approach is suggested by
examining the $(L+1)$-particle cuts of an $L$-loop amplitude.  Using
such a cut we can decompose the loop integrals into products of tree
diagrams which then simplifies the bookkeeping.\footnote{At six
loops and beyond it turns out that there are pseudo-conformal integrals
which do not have an $(L+1)$-particle cut. These can however be
obtained from a ``parent diagram'' containing an $(L+1)$-particle
cut, by canceling one of the propagators.} Our procedure will
be:
\begin{enumerate}
\item Construct the set of all possible amputated tree configurations
on each side of the cut.
\item Identify all possible loop integrals by sewing each configuration
from the left side of the cut with each configuration on the right
side of the cut.
\item Identify all possible overall factors in each integral
which make it conformal.
\end{enumerate}
Because the conformal properties are most obvious in the dual
representation described in \sect{ConformalIntegralsSection}, we have
found it convenient to work with dual coordinates and translate back
to the momentum representation at the end.

\subsection{Constructing all dual tree diagrams}

\begin{figure}[t]
\centerline{\epsfxsize 4.5 truein \epsfbox{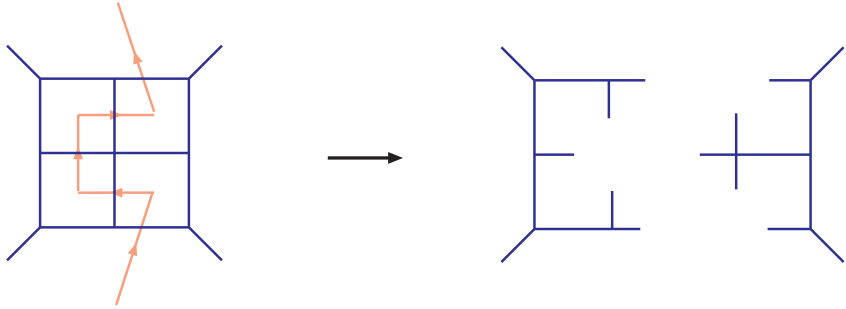}}
\caption{The four-loop ``window'' diagram.  The lighter colored line
running through the diagrams is a five-particle cut which separates
the diagram into a product of tree diagrams.}
\label{CutWindowFigure}
\end{figure}

\begin{figure}[t]
\centerline{\epsfxsize 5.5 truein \epsfbox{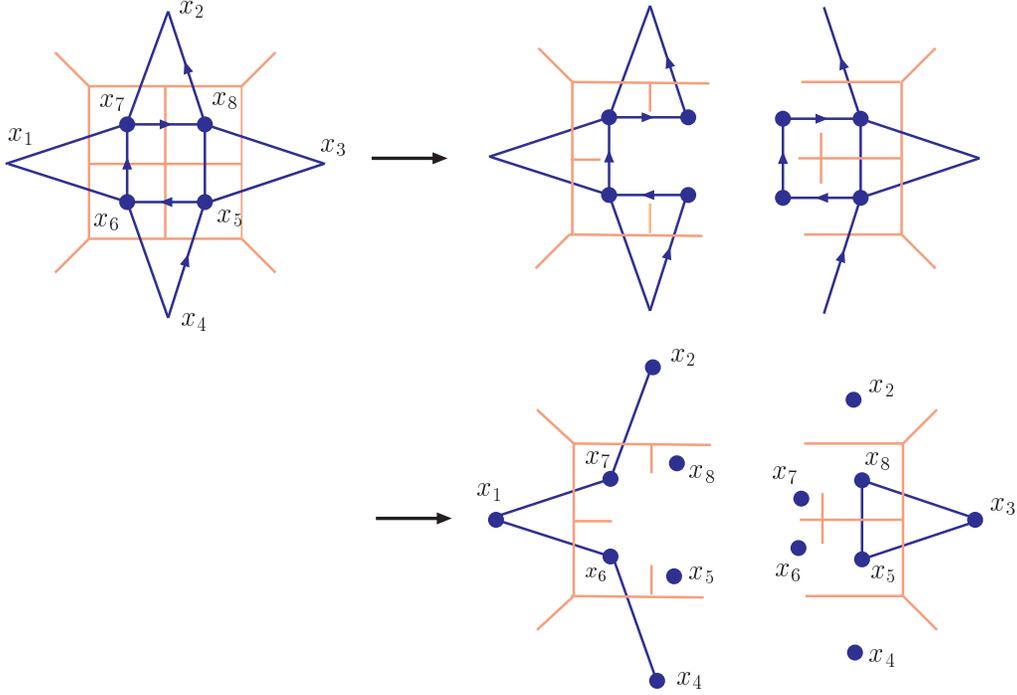}}
\caption{The dual diagram corresponding to the window diagram.
The lines with the arrows indicate the cut which separates the
dual diagram into a product of tree dual diagrams.}
\label{DualWindowFigure}
\end{figure}

It is useful to consider first the reverse procedure of splitting an
$L$-loop integral into two trees via an $(L+1)$-particle cut.  As an
illustration, consider the four-loop ``window'' integral as cut in
\fig{CutWindowFigure}.  This integral contributes to the four-loop
four-point amplitude~\cite{BCDKS}.  As shown in the figure, a
five-particle cut separates the integral into a product of two tree
diagrams.  In dual space, we also split integrals into tree diagrams
with a five-particle cut.  \Fig{DualWindowFigure} depicts the
separation of the corresponding dual diagram into tree diagrams, which
we carry out in two steps.  In the first step of the figure we divide
the dual diagram along the cut marked with arrows, keeping the cut
line on both sides.  In the next step we drop the lines with
arrows, giving us dual diagrams corresponding to amputated ({\it i.e.}
with external propagators removed) tree-level momentum-space diagrams.
In this representation each dual line crosses an internal propagator
of the momentum-space diagrams.  The diagrams will always 
have a fixed cyclic ordering. The
dual-diagram points also respect this cyclic ordering.  In this
example, for the tree amplitude on the left, the points are ordered
$\{x_4, x_1, x_2, x_8, x_7, x_6, x_5 \}$, while the points for the
tree diagram on the right are ordered $\{ x_2, x_3, x_4, x_5, x_6,
x_7, x_8 \}$.

Our systematic construction of all dual loop diagrams simply
reverses the process in this example.
At $L$ loops we start with two ordered lists of $L+3$
points: $\{x_1,x_2,x_{L+4}, x_{L+3}, \ldots, x_5,x_4 \}$ for the left tree
and $\{x_2, x_3, x_4, x_5, \ldots, x_{L+4} \}$ for the right tree.  These
lists correspond to $(L+1)$-particle cuts in the $s_{12}$ channel; the
corresponding construction in the $s_{23}$ channel is easily obtained
by relabeling the final result.  The assignment of labels 
$x_1, x_2,x_3,x_4$ to points is determined by the external momenta, and the 
remainder follow from the cyclic ordering.

We obtain all possible pairs of dual tree diagrams by connecting
non-adjacent points with $1/x_{ij}^2$ propagators in all possible
ways such that the lines do not cross. Dual diagrams where
nearest-neighbor points are connected are not included as they correspond
to momentum-space diagrams whose external propagators have not been
truncated.

After identifying the possible dual tree diagrams we restore the dual
lines representing the cut, by retracing our steps in the example
shown in \fig{DualWindowFigure}.  That is, in both sets of tree
diagrams we draw lines with arrows connecting $x_4$ to $x_5$, {\it
etc.}, to $x_{L+4}$, which is then connected to $x_2$.  Once this is
done the pairs of tree diagrams can be glued together along the lines
with arrows, which then gives us the $L$-loop dual diagrams that we wish
to construct. At this stage we can remove diagrams trivially related
by cyclic or flip symmetry.

\subsection{Finding the pseudo-conformal integrals}

Once we have a set of candidate loop-level dual diagrams, we
must find the numerator factors necessary to make the corresponding
integrals conformal.  This can be accomplished as follows~\cite{DHSS,BCDKS}:
\begin{itemize}
\item If one of the internal points $\{x_5, x_6, \ldots, x_{L+4} \}$
appears in less than four dual propagators, discard the
diagram as it cannot be made conformal.

\item To determine possible numerator factors one first identifies
all external points from the set $\{x_1,x_2,x_3, x_4\}$ appearing in one
or more dual propagators, and all internal points in the set $\{x_5,
x_6, \ldots, x_{L+4} \}$ appearing in five or more dual propagators.
All such points require numerator factors $x_{ij}^2$ to cancel the
extra conformal weight.  That is, the number of times a given external $x_i$
appears in the dual propagators minus the number of times it appears in
the numerator should be zero.  Similarly, for an internal point $x_i$,
the number of times it appears in the dual propagators minus the
number of times it appears in the numerator should be four.  To find
the conformally invariant integrals we sweep through products of all
candidate numerators $x_{ij}^2$ to identify the ones where the conformal
invariance constraints are satisfied.  (In principle, there might also
be an overall resulting factor of $1/s = 1/x_{24}^2$ or $1/t =
1/x_{13}^2$, but this does not occur at five loops, nor do we expect
such contributions to enter the amplitudes with non-zero coefficients
at any loop order.)

\item If the previous step yields a previously-identified
pseudo-conformal integral, go on to the next case.  Such repeated
integrals can arise when a numerator factor cancels a propagator or
when diagrams are related by symmetries.

\end{itemize}

Once we have the set of conformal dual diagrams we can convert
these back to momentum space with a change of variables,
\begin{eqnarray}
&&
k_1 = x_{41}\,, \hskip 1 cm
k_2 = x_{12}\,, \hskip 1 cm
k_3 = x_{23}\,, \hskip 1 cm
k_4 = x_{34}\,, \nn\\
&&
l_1 = x_{45}\,, \hskip 1 cm
l_2 = x_{56}\,, \hskip 1 cm
\ldots \,, \hskip 1 cm
l_{L+1} = x_{(L+4)2}\,,
\end{eqnarray}
where the $l_i$ are the momenta of the lines in $(L+1)$-particle cut
used in the construction.  Since our construction was only a
bookkeeping device for finding pseudo-conformal integrals, at the end there
is no on-shell restriction on the $l_i$. 

In the next section we apply this procedure to construct a basis of all
pseudo-conformal integrals appearing in the five-loop planar MSYM
amplitudes.

\section{The five-loop planar pseudo-conformal integrals}
\label{ResultSection}

\subsection{The five-loop pseudo-conformal integral basis}

\begin{figure}[t]
\centerline{\epsfxsize 7 truein \epsfbox{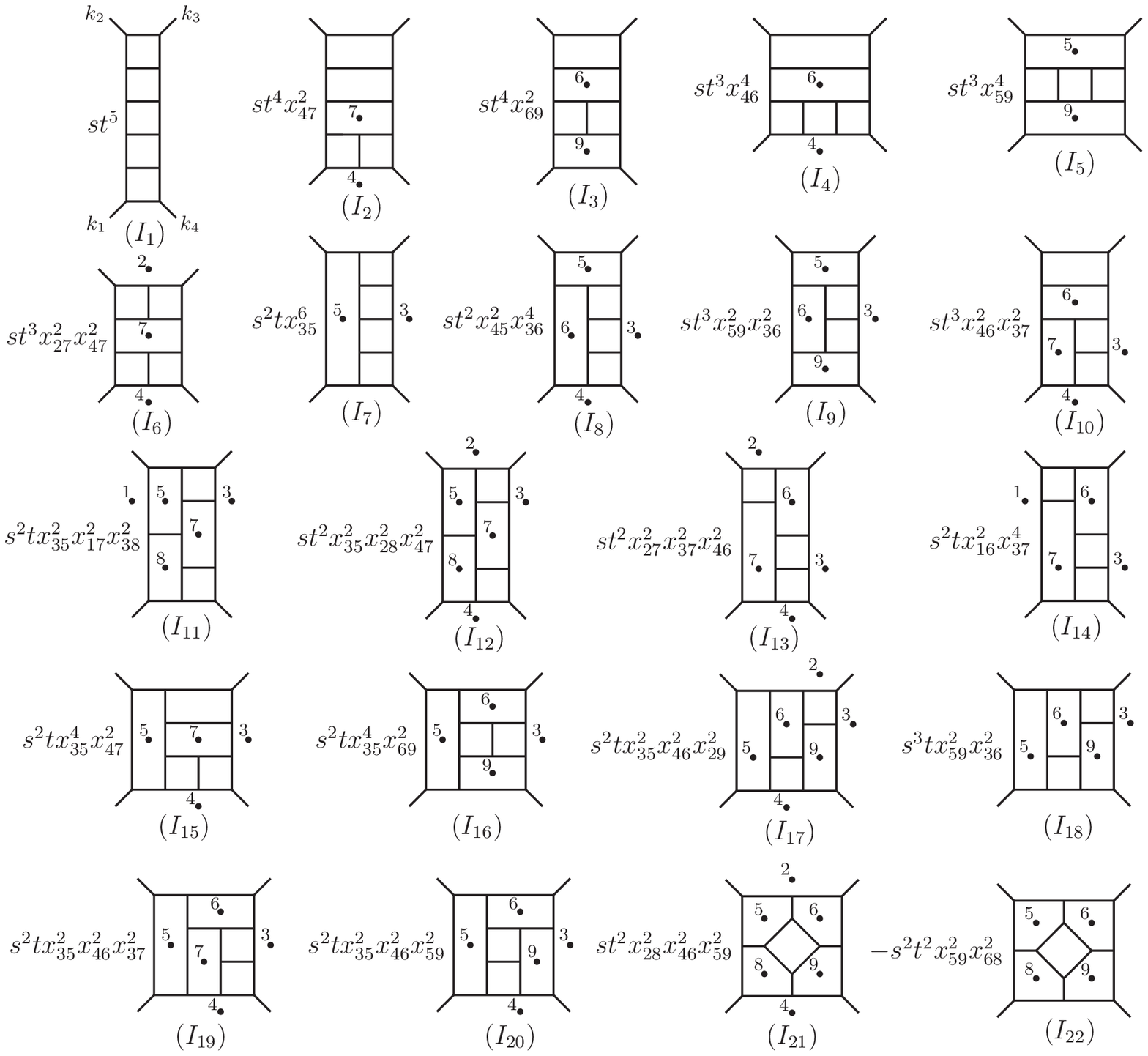}} \caption{All
pseudo-conformal integrals with only cubic vertices that contribute to the
amplitude.  The relative signs are determined from unitarity cuts in
\sects{MaximalCutSection}{ConfirmingCutsSection}.}
\label{cubicsFigure}
\end{figure}

\begin{figure}[t]
\centerline{\epsfxsize 7. truein \epsfbox{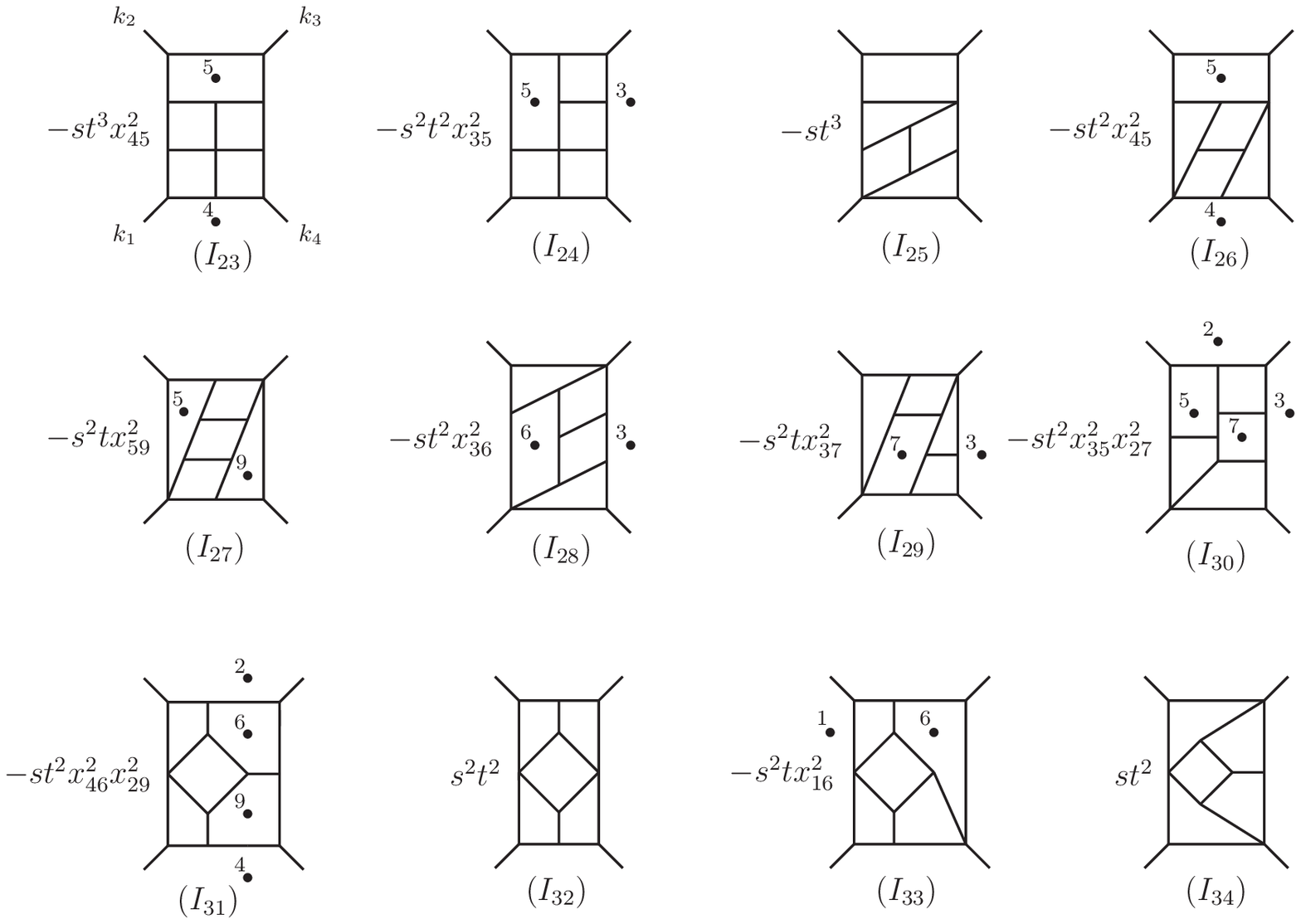}}
\caption{All pseudo-conformal integrals with cubic and quartic vertices that
contribute to the amplitude. The relative signs are determined from
unitarity cuts in \sects{MaximalCutSection}{ConfirmingCutsSection}.}
\label{quarticsFigure}
\end{figure}

\begin{figure}[t]
\centerline{\epsfxsize 6.5 truein \epsfbox{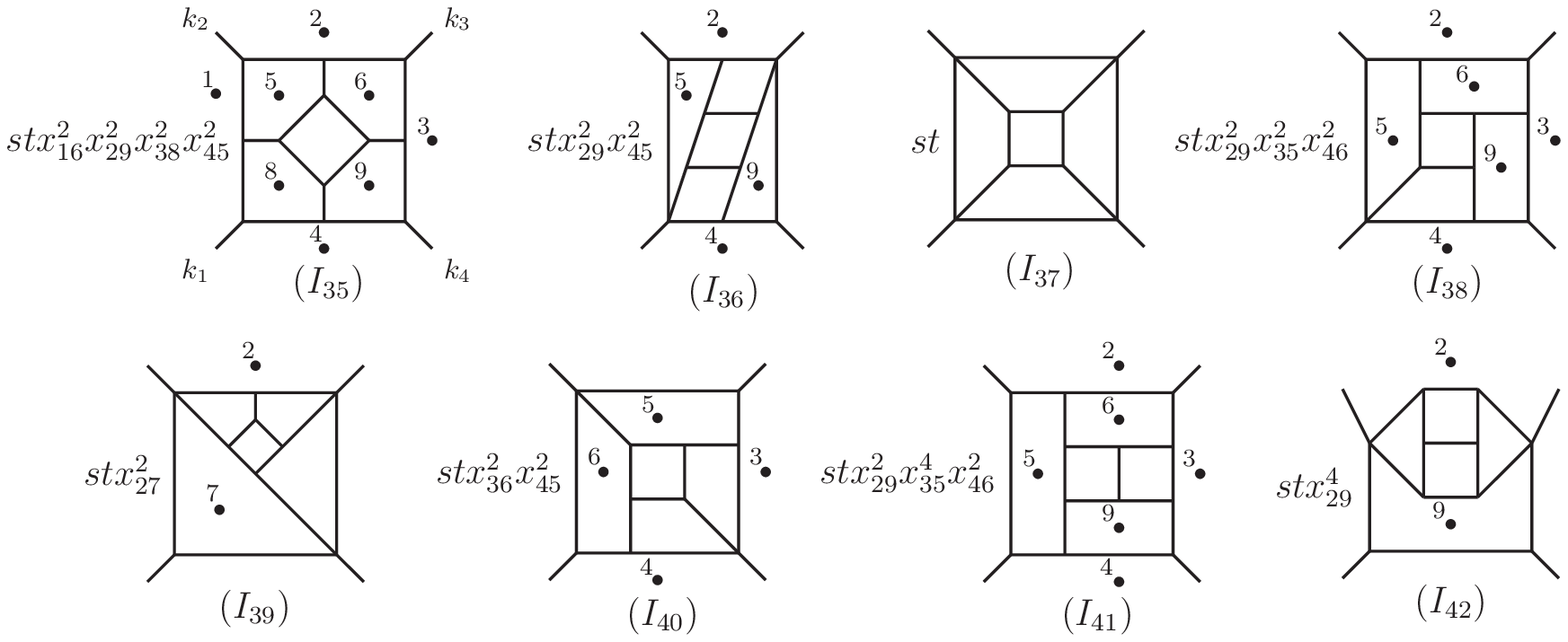}}
\caption{A class of pseudo-conformal integrals which do not contribute to
the amplitude, as determined from the unitarity cuts. All these have
exactly one factor of $st$.} \label{STIntegralsFigure}
\end{figure}

\begin{figure}[t]
\centerline{\epsfxsize 6.5 truein \epsfbox{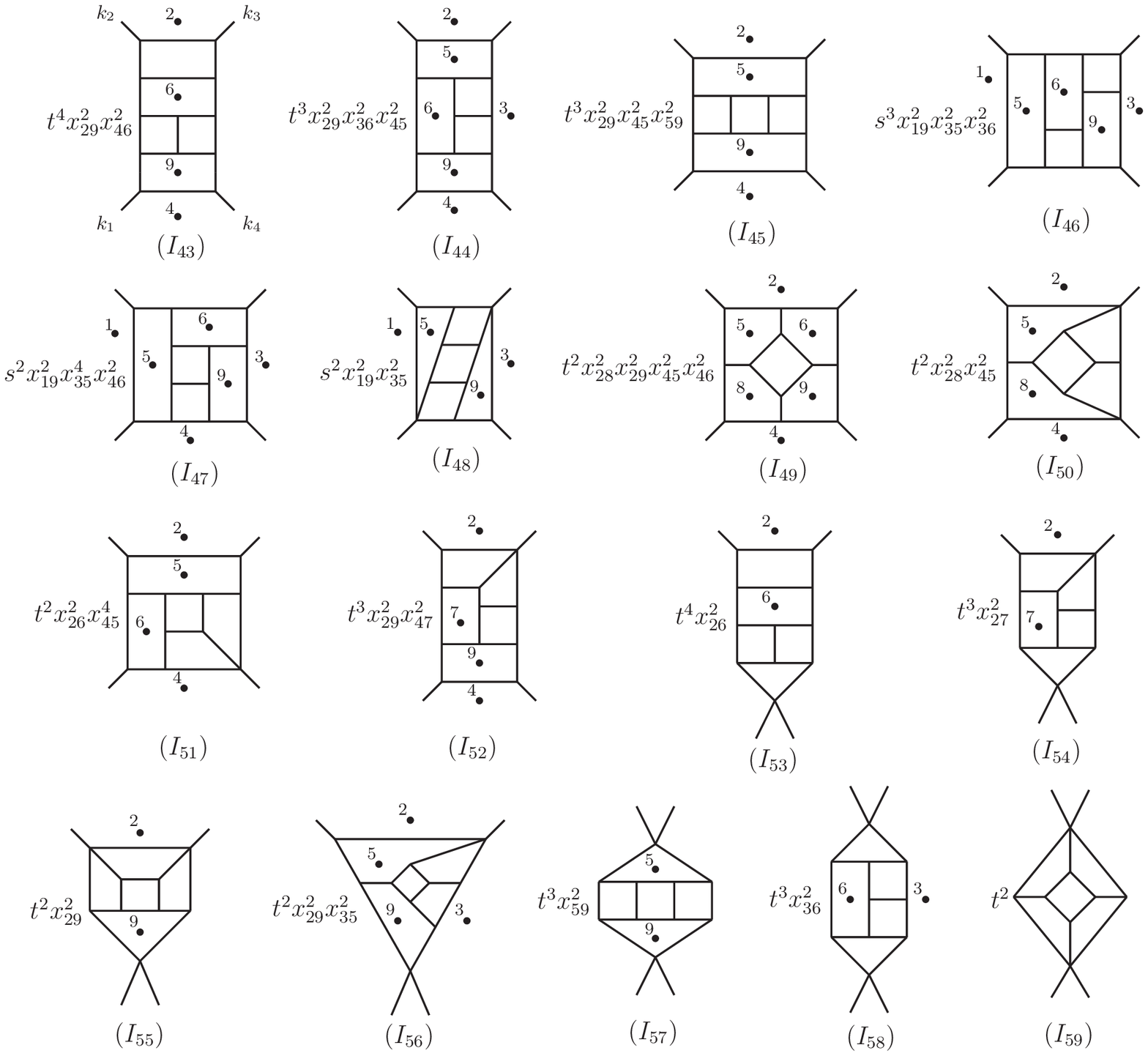}} \caption{The
non-$s t$ class of pseudo-conformal integrals.  They do not contribute to
the amplitude.}
\label{nonSTIntegralsFigure}
\end{figure}

Following the procedure described in the previous section we find a
total of 59 independent pseudo-conformal integrals potentially
present in the five-loop
four-point planar amplitude (not counting those related by
permutations of external legs).  They are shown in
figs.~\ref{cubicsFigure}, \ref{quarticsFigure},
\ref{STIntegralsFigure} and \ref{nonSTIntegralsFigure}.  The `parent'
integrals, shown in \fig{cubicsFigure}, have only cubic vertices.  The
remaining integrals have both cubic and quartic vertices.
They may be obtained by omitting propagators
and modifying numerator factors present in the parent integrals,
As we shall show in the following section using unitarity cuts, the integrals
in \figs{cubicsFigure}{quarticsFigure} appear in the amplitude
(\ref{LoopOverTree}) with relative coefficients of $\pm 1$, which we have
absorbed into the definitions of the numerator factors in the figures.
The remaining ones shown in \figs{STIntegralsFigure}{nonSTIntegralsFigure}
do not appear at all.  We do not have an explanation for the remarkable
simplicity of the coefficients of the integrals, but presumably it is
tied to the superconformal invariance of the theory.

\begin{figure}
\centerline{\epsfxsize 3.5 truein \epsfbox{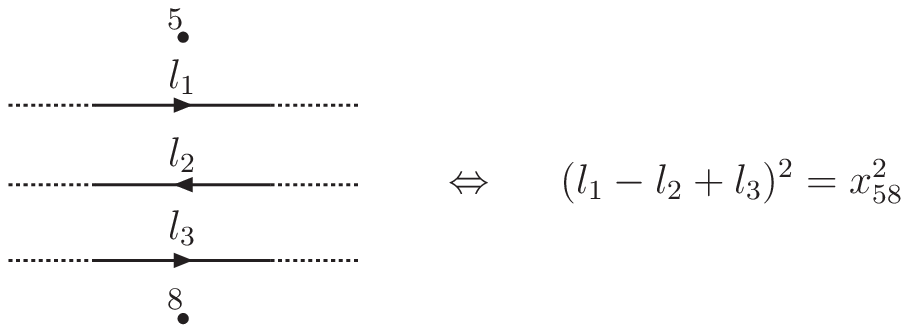}}
\caption{The notation used in this section for listing out the
pseudo-conformal integrals contributing at five loops. The momentum flow
through a line connecting points 5 and 8 gives the momentum
invariant $x^2_{58}$. } \label{dconventions}
\end{figure}

We draw the diagrams in momentum space, but also include the relevant
$x_i$ for tracking numerator factors.  The numerators are written out
as Mandelstam variables $s = (k_1 +k_2)^2$ and $t=(k_2+k_3)^2$ or as
dual invariants, $x_{ij}^2$ . As discussed in
\sect{ConformalIntegralsSection}
a dual invariant $x^2_{ij}$ is equal to $K^2$
where $K$ is the total momentum flowing through a line spanned between
points $i$ and $j$.  For example, in  \fig{dconventions},
$K^2 = (l_1 - l_2 + l_3)^2 = x_{58}^2$.

Some of the integrals have identical sets of propagators but differing
numerator factors.  These {\it sibling\/} integrals would be identical
were one to omit the numerators.
 Examples are $I_{11}$ and $I_{12}$ or
$I_{21}$ and $I_{22}$ in \fig{cubicsFigure}. The numerator
factors often have different symmetries than the propagators in
any given integral.  The different numerator factors in
sibling integrals will also typically have different symmetries.  For
example, integral $I_{22}$ is completely symmetric under a cyclic
permutation of its arguments, $\{1,2,3,4\}\rightarrow \{2,3,4,1\}$
(corresponding to a $\pi/2$ rotation of the diagram in~\fig{cubicsFigure})
and under flips, $\{1,2,3,4\}\leftrightarrow\{4,3,2,1\}$ (corresponding
to reflection of the diagram).  Its sibling $I_{21}$, in contrast, has only one
symmetry, $\{1,2,3,4\}\rightarrow \{3,4,1,2\}$ 
(corresponding to a rotation of the diagram by $\pi$ radians). Accordingly,
$I_{22}$ appears only once in the amplitude, but
$I_{21}$ appears four times.  This makes it inconvenient
to combine them into a single integral.

All planar five-loop pseudo-conformal integrals with the exception of
$I_{55}$ have at most four-point
vertices.  ($I_{55}$ has quintic
vertices where the external legs attach.
Internal quintic vertices do not occur in conformal
integrals until seven loops.)

\subsection{The five-loop four-point amplitude}

In \sects{MaximalCutSection}{ConfirmingCutsSection}, we evaluate a
sufficient number of cuts in order to determine the numerical
prefactors of each pseudo-conformal integral as it appears in the
amplitude~(\ref{LoopOverTree}).
We find that the complete five-loop four-point MSYM planar amplitude is,
\begin{eqnarray}
M_4^\fiveloop(1,2,3,4) &=&
-{1\over 32}
\Bigl[ \Bigl(I_1 + 2 I_2 + 2 I_3 + 2I_4 + I_5 + I_6 + 2 I_7 + 4 I_8 + 2 I_9 +
4 I_{10} \nn \\
&& \null \hskip 1 cm
+ 2 I_{11} + 4 I_{12} + 4 I_{13} + 4 I_{14}
+ 4 I_{15} + 2 I_{16} + 4 I_{17} + 4 I_{18} + 4 I_{19} + 4 I_{20}  \nn \\
&& \null \hskip 1 cm
+ 2 I_{21} + 2 I_{23} + 4 I_{24} + 4 I_{25} + 4 I_{26}
+ 2 I_{27} + 4 I_{28} + 4 I_{29} + 4 I_{30}  \nn \\
&& \null \hskip 1 cm
+ 2 I_{31} + I_{32} + 4 I_{33} + 2 I_{34}
+ \{s \leftrightarrow t \} \Bigr)
+ I_{22}  \Bigr] \,,
\label{FiveLoopAnsatz}
\end{eqnarray}
where the integrals are shown in \figs{cubicsFigure}{quarticsFigure},
and $M_4^\fiveloop$ is defined in \eqn{LoopOverTree}.  There are a
total of 193 integrals in the sum. As the integrals depend only on the
kinematic invariants $s$ and $t$, instead of having leg labels, each integral
can appear only as $I_j(s,t)$ or as $I_j(t,s)$. In
\eqn{FiveLoopAnsatz}, we have suppressed the arguments ``$(s,t)$'' and
combined identical terms, leaving a symmetry factor in front.  The
relative signs between integrals, determined from the unitarity
cuts in \sects{MaximalCutSection}{ConfirmingCutsSection}, have been
incorporated in the numerator factors in
\figs{cubicsFigure}{quarticsFigure}, though we have chosen to leave an
overall sign outside the integrals.  The normalization factor of
$1/32$ follows the conventions of refs.~\cite{BDS,BCDKS} and accounts
for the factor of $2^L$ in \eqn{LeadingColorDecomposition}.  
The integrals in \eqn{FiveLoopAnsatz}
are therefore normalized as:
\begin{equation}
(-i e^{\epsilon \gamma} \pi^{-D/2})^5 
\int  \Bigl( \prod_{i=1}^5 d^D l_i \Bigr) \,  {N \over \prod_j p_j^2}
\label{IntegralNormalization}
\end{equation}
where the $l_i$ are five independent loop momenta, $N$ is the numerator
factor appearing as the coefficient of the diagrams given in 
figs.~\ref{cubicsFigure}-\ref{nonSTIntegralsFigure}, and the $p_j^2$
correspond to the propagators of the diagrams.

To understand the relative signs of the diagrams we classify terms into
those derived from the rung rule and the rest.
Any diagram generated by the rung rule in \fig{RungRuleFigure}
simply inherits the sign of the lower loop diagram from which it was
derived.
This gives the correct numerator factor, including the sign, for all
contributing integrals containing only cubic vertices, except for
$I_{22}$ --  the only diagram in \fig{cubicsFigure} having a
non-rung rule numerator. Integrals with quartic vertices, such as
$I_{24}, I_{27}, I_{28}$ and $I_{29}$, are given by the rung rule
applied to known~\cite{BCDKS} four-loop diagrams. Using two-particle
cuts, other examples of integrals whose prefactors are easy to
understand are $I_{23}, I_{25}, I_{26}, I_{41}$ and $I_{42}$. The
latter two have vanishing coefficient because their four-loop parent
diagrams also have vanishing coefficients. Integral $I_{33}$
can be understood in terms of a rung inserted between an external
leg and an internal line.

\begin{figure}[t]
\centerline{\epsfxsize 4 truein \epsfbox{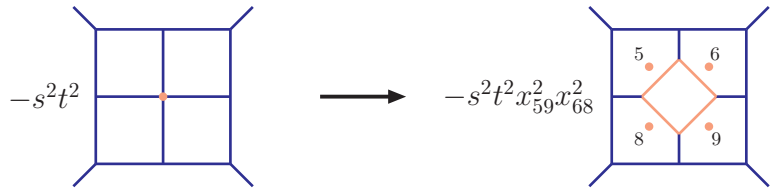}} \caption{The
``substitution rule'' is a rule for replacing any four-point vertex
with terms in a four-point amplitude.  Here integral $I_{22}$ is
obtained by substituting the pseudo-conformal box integral appearing
in the one-loop amplitude into the central vertex of the four-loop
``window'' diagram.} \label{subruleFigure}
\end{figure}

Another class of prefactors and signs can be understood from a
``substitution rule''.  Consider diagram $I_{22}$ in
\fig{cubicsFigure}.  As shown in \fig{subruleFigure}, this diagram
inherits its prefactor and sign by replacing the four-point vertex
by a one-loop box integral. (The one-loop
box enters with a relative plus sign.) The numerator
factor $x_{59}^2 x_{68}^2$ of the substituted box is simply the factor
needed to make the box conformal. 
The negative sign is inherited from the
sign of the four-loop diagram. Also other signs can be understood from
this substitution rule. For example, the sign on $I_{30}$ and the zero
coefficient on $I_{53}$ follow from similar substitutions on four loop
conformal diagrams. This rule can more generally be understood as a
substitution of the normalized four-point function into a four-point
vertex, which can obtained using generalized cuts.

With the rung rule, two-particle cuts, and substitution rule we may
understand the signs of all diagrams that appear in the amplitude,
except for $I_{31}$, $I_{32}$ and $I_{34}$. As of yet we have not
found a rule giving the sign of these diagrams, other than resorting
to computations of cuts. 

At four loops~\cite{BCDKS}, integrals not containing at least one
factor of $s$ and also one factor of $t$ are absent from the
amplitude.  This is again true at five loops.  Factorization arguments
using complex momenta can give a suggestive explanation of this
property.  As we already noted, supersymmetry identities ensure that
after dividing by the tree amplitude, the non-vanishing MSYM
four-point loop amplitudes are identical for all external helicity and
particle configurations.  Consider then the helicity configuration
$1^-,2^+,3^-,4^+$, with all external legs gluons.  The tree amplitude
\begin{equation}
A_4^\tree(1^-,2^+,3^-,4^+)= i {\spa1.3^4 \over \spa1.2 \spa2.3
                               \spa3.4 \spa4.1} = -i {\spa1.3^2 \spb2.4^2
                                       \over s t}\,,
\end{equation}
factorizes in both the $s$ and $t$ channels into products of
three-point vertices. The absence of compensating factors of $s$ and
$t$ would imply factorization into one-particle irreducible loop three
vertices.  Such vertices have not appeared in the factorization of any
previous MSYM amplitude, and so it is not surprising that
the offending integrals, shown in \fig{nonSTIntegralsFigure}, do not
contribute here either.

We may understand the remaining vanishing coefficients using the known
harmonic-superspace power counting of MSYM~\cite{HoweStelleNew}.  It
is compatible in $D=4$ with six powers of loop momenta canceling from
integral numerators.  The missing engineering dimensions of the amplitude
are then supplied by external momenta,
requiring at least three powers of either $s$ or
$t$.
 This result agrees with the arguments of
refs.~\cite{BRY,BDDPR}, which provide a bound on dimensions for which the
$L$-loop amplitude is ultraviolet finite,
\begin{equation}
D< {6\over L} + 4 \,,  \hskip 2cm (L>1)\,.
\label{Finiteness}
\end{equation}
It is natural to assume that this power counting holds independently
for each integral.  This rules out integrals which do not have
at least three powers of $s$ or $t$. This includes all of those in
\fig{STIntegralsFigure} and those with either a single power of $s^2$
or $t^2$ in \fig{nonSTIntegralsFigure}.

\subsection{All-loop structure}

Inspecting the contributions of the integrals in the basis
 to the five-loop four-point
amplitude reveals the following features:

\begin{itemize}

\item All pseudo-conformal integrals containing a factor of $s^2 t$ or
$t^2 s$ (and possibly additional powers of $s$ or $t$)
enter the amplitude with relative weight of $+1$ or $-1$.

\item Any pseudo-conformal integral without a factor of $s^2 t$ or $t^2 s$
has a vanishing coefficient.

\item All integrals that could be obtained from the rung rule
or from two-particle cuts
inherit their weights from the lower-loop integrals used to construct them.

\item All contributing integrals satisfy the ultraviolet
finiteness bound (\ref{Finiteness}).
\end{itemize}

\begin{figure}[t]
\centerline{\epsfxsize 2 truein \epsfbox{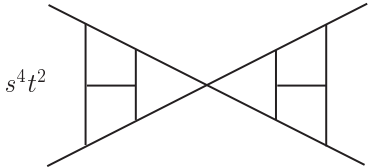}} \caption{
A pseudo-conformal integral with a vanishing coefficient
in the six-loop amplitude.} \label{Conformal6ExampleFigure}
\end{figure}

These observations seems to suggest that, in general, the set of
pseudo-conformal integrals with nonvanishing coefficients are the ones with
a prefactor divisible by either $s^2 t$ or $t^2 s$. However, at six
loops a new structure appears where a pseudo-conformal integral is a simple
product of lower-loop integrals, as displayed in
\fig{Conformal6ExampleFigure}.  We have checked that the conformal
integral in \fig{Conformal6ExampleFigure} does not contribute to the
amplitude, although its prefactor is divisible by $s^2 t$.  While
this particular integral does not appear in the amplitude,
its existence suggests that at higher loops there will be additional classes of
pseudo-conformal integrals with vanishing coefficients.

\section{Generalized Unitarity}
\label{UnitaritySection}

The unitarity method~\cite{NeqFourOneLoop, Fusing, 
DDimUnitarity,OneLoopReview,DimShift, GeneralizedUnitarity,
BCFUnitarity} has proven an effective means for computing
scattering amplitudes in gauge and gravity theories.
So-called generalized unitarity is particularly powerful
for computing amplitudes~\cite{GeneralizedUnitarity,
BCFUnitarity}, as it allows an $L$-loop amplitude to be built directly
from products of tree amplitudes.  When combined with complex
momenta~\cite{WittenTopologicalString,BCFUnitarity,BCFW}, it
allows the use of maximal cuts, in which {\it all\/} propagators in
an integral are cut.
(The term ``generalized unitarity'', corresponding
to leading discontinuities of diagrams, dates back to
ref.~\cite{EarlyGeneralizedUnitarity}.)

We begin our discussion with a brief review, including
earlier applications of maximal cuts to the computation of
two-loop amplitudes~\cite{BCFUnitarity, BuchbinderCachazo}.
We record a number
of observations useful for computation at higher loops.  In
\sect{MaximalCutsSubsection} we modify the maximal-cut procedure
and use it to determine
the coefficients of all pseudo-conformal integrals appearing in the MSYM
five-loop four-point amplitudes efficiently and systematically.

\subsection{Maximal cuts}
\label{MaximalCutsSubsection}

Cut calculations can be simplified by increasing the number of cut
legs. This isolates a smaller number of integrals, making it simpler
to determine the values of their coefficients.  This technique is
especially powerful for computing one-loop MSYM amplitudes, because
only box integrals can appear~\cite{NeqFourOneLoop}.  As observed by
Britto, Cachazo and Feng~\cite{BCFUnitarity}, taking a quadruple cut,
where all four propagators in a box integral are cut, freezes the
four-dimensional loop integration.  This allows its kinematic
coefficient to be determined {\it algebraically\/}, with no
integration (or integral reduction) required.  The use of complex
momenta, as suggested by twistor space
theories~\cite{WittenTopologicalString}, makes it possible to define
massless three vertices and thereby to use quadruple cuts to determine
the coefficients of all box integrals including those with massless
external legs.

For three massless momenta $k_a$, $k_b$ and $k_c=-(k_a+k_b)$ one has
the following consistency requirement
\begin{equation}
0 = k_c^2 =(k_a+k_b)^2=  2k_a \cdot k_b = \spa{a}.{b} \spb{b}.a \,.
\label{oscond}
\end{equation}
For real momenta in Minkowski signature, $\lambda_{k_a}$ and
${\tlambda}_{k_a}$ (see \eqn{lambdadef}) are complex conjugates of
each other (up to a sign determined by incoming or outgoing nature of
the corresponding particle).  Hence if $\spa{a}.b$ vanishes then
$\spb{a}.b$ must also vanish. This constraint holds for 
all three legs $a,b,c$,
leaving no non-vanishing quantities out of which to build a three
vertex.  If the momenta are taken to be complex, however, the two
spinors $\lambda_{k_a}$ and ${\tlambda}_{k_a}$ are independent. This
gives two independent solutions to \eqn{oscond},
\begin{equation}
\spa{a}.b=0\,, \hskip 1 cm   \hbox {\it or} \hskip 1 cm 
\spb{a}.b=0 \,,
\label{TwoBranches}
\end{equation}
with the other spinor product non-vanishing in each case.
In the three-gluon case, there are overall two possible solutions:
all $\lambda$s proportional, and hence all $\spa{i}.{j}$ vanishing,
or all $\tlambda$ proportional, and hence all $\spb{i}.j$ vanishing.
This means that exactly one of,
\def\VT#1{A_3^{(#1)}}
\begin{eqnarray}
\VT- \equiv A_3^{\tree}(a^-,b^+,c^+)=-i
 \frac{ \spb{b}.{c}^3}{ \spb{a}.{b} \spb{c}.a} \, ,
\label{ThreeVertexMinus}  \\
\VT+ \equiv A_3^{\tree}(a^+,b^-,c^-)=
     i \frac{\spa{b}.c^3}{\spa{a}.b \spa{c}.a} \,,
\label{ThreeVertexPlus}
\end{eqnarray}
does not vanish.  Similar statements hold for amplitudes involving fermions or
scalars: one of the two independent helicity configurations
will not vanish.  The non-vanishing amplitudes involving a fermion 
pair are, 
\begin{eqnarray}
A_3^{\tree}(a^-_{\! f},b^+_{\! f},c^+)= -i
 \frac{ \spb{b}.{c}^2 }{ \spb{a}.{b}} \, ,
\label{ThreeVertexMinusFermion}  \\
A_3^{\tree}(a^+_{\! f},b^-_{\! f},c^-)=
     -i \frac{\spa{b}.c^2 }{\spa{a}.b } \,,
\label{ThreeVertexPlusFermion}
\end{eqnarray}
where the subscript $f$ denotes a fermionic leg.
Similarly the non-vanishing scalar amplitudes are,
\begin{eqnarray}
A_3^{\tree}(a^-_{s},b^+_{s},c^+)= - i
 \frac{ \spb{b}.{c} \spb{c}.a }{ \spb{a}.{b}} \, ,
\label{ThreeVertexMinusScalar}  \\
A_3^{\tree}(a^+_{s},b^-_{s},c^-)=
     i \frac{\spa{b}.c \spa{c}.{a}}{\spa{a}.b } \,,
\label{ThreeVertexPlusScalar}
\end{eqnarray}
where the subscript $s$ denotes a scalar leg.  (For complex
scalars, the two helicities correspond to particle and antiparticle.)
We have chosen the signs in these amplitudes to be consistent with the
supersymmetry Ward identities~\cite{SWI}, as given in
ref.~\cite{Fusing}, and with the parity conjugation rules of
ref.~\cite{QQGGG}.

The method of quadruple cuts has been generalized by Buchbinder and
Cachazo~\cite{BuchbinderCachazo} to two loops using hepta- and
octa-cuts. Although the two-loop four-point double-box integral only
has seven propagators it secretly enforces
an additional, eighth
constraint, yielding an octa-cut which localizes the integration
completely.  But as the authors of ref.~\cite{BuchbinderCachazo}
point out this last cut condition is not really necessary for 
the evaluation of the two-loop four-point
amplitude: the integrand was already independent of loop momenta
after imposing the constraints from the seven on-shell propagators in
the hepta-cut.  We will return to this point below.

\begin{figure}
\centerline{\epsfxsize 5 truein \epsfbox{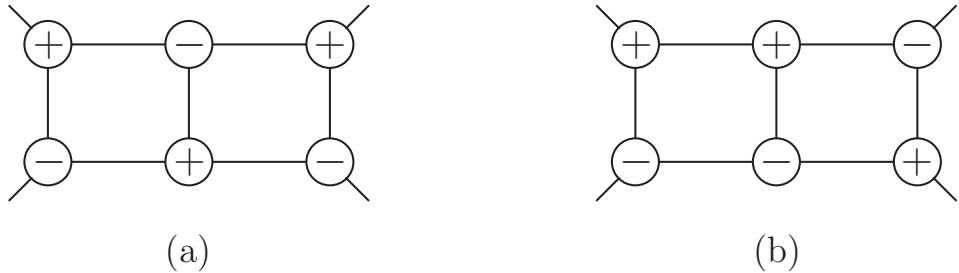}} \caption{A
pictorial representation of two kinematic solutions to the
four-point hepta-cut equations. The diagrams representing
the remaining four solutions can be
obtained by reflection symmetries of these. A `\PlusVertex'  vertex
represents a three-point tree amplitude involving only $\lambda$
spinors and a `\MinusVertex'  vertex represents an amplitude
involving only $\widetilde\lambda$ spinors. 
All lines are cut and
carry on-shell momenta.} \label{heptacutFigure}
\end{figure}

Let us then focus on the constraints imposed by the delta functions
corresponding to the propagators alone.
In the hepta-cut construction of ref.~\cite{BuchbinderCachazo},
various classes of solutions are allowed by the seven delta function
constraints arising from localizing the propagators.  In the context
of generalized unitarity these delta functions correspond to solving
the on-shell cut conditions $l_i^2=0$. As discussed above the solution
to these conditions is always complex when three-point vertices are
present.  These cut conditions have a discrete set of solutions
because of the two-fold choice in \eqn{TwoBranches}
at each three-point vertex. 
Each of these solutions depends on continuous parameters,
corresponding to the degrees of freedom not frozen by the cut conditions.
The discrete choice coincides with the choice of
three-point amplitude at each vertex $\VT{\pm}$ as given in
\eqns{ThreeVertexMinus}{ThreeVertexPlus} (or similar vertices for
fermionic and scalar lines), which suggests a convenient way to represent the
possible solutions using additional labels at the vertices of
the cut diagrams. 
For the four-point hepta-cut two
inequivalent arrangements of three-point vertices are shown in
\fig{heptacutFigure}. External legs represent outgoing external momenta while
internal lines represent cut propagators and thus on-shell loop momenta. 
The signs inside the blobs in the diagrams indicate the corresponding
choice for the three-point vertex, and implicitly, that the spinors
of the opposite helicity are proportional.  A `\MinusVertex' vertex
will  have all $\lambda$ spinors proportional to each other, so that
the vertex is built out of $\tlambda$ spinors of the attached legs,
while the roles of the two kinds of spinors are interchanged for a
`\PlusVertex' vertex. We can sum over all possible solutions to
obtain the multiply-cut integrand, as was done in
ref.~\cite{BuchbinderCachazo}.  For the sevenfold cut of the
double-box diagram, there are six distinct solutions of the 
two types shown in \fig{heptacutFigure}.

\begin{figure}
\centerline{\epsfxsize 5.2 truein \epsfbox{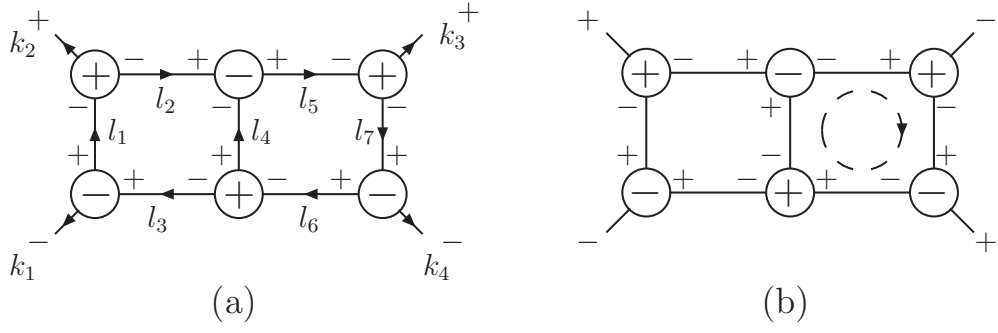}}
\caption{A singlet hepta-cut (a) and one of the two helicity
configurations (b) in the simplest non-singlet cut.  The latter allows
gluons, fermions and scalars to propagate in the loop indicated by a
dashed circle.  The other configuration is obtained by flipping all
the helicity signs of the legs in this loop.  All lines in this figure
are cut and carry on-shell momenta. The arrows in (a) refer to the direction
of momentum flow.}
\label{singlet}
\end{figure}

Once we choose external helicities, as in \fig{singlet}, the blobs
then dictate the possible assignments of helicities for internal
lines. The rules for finding the complete set of kinematic solutions
associated with a given assignment of plus and minus labels
to a diagram are as follows:
\begin{itemize}

\item A `\MinusVertex' label means the three $\lambda$ spinors
corresponding to the lines attached to the blob are proportional to each
other.  Similarly, a `\PlusVertex' label denotes having the three $\tlambda$
spinors proportional to each other.

\item If one of the lines attached to a vertex 
is an external line $k_i$ then the
spinors are proportional to an external spinor, either
$\lambda_{k_i}$ or $\tlambda_{k_i}$.

\item If a `\PlusVertex' vertex is directly connected to another
`\PlusVertex' vertex
then all $\tlambda$s of the lines attached to both vertices are
proportional to each other.  A similar statement holds for the
$\lambda$s of two connected `\MinusVertex' vertices.

\item If there is a chain of vertices of the same sign connecting any
two external lines then the diagram vanishes, because
one cannot solve the on-shell and momentum
conservation constraints for the diagram. A solution would require
that two {\it external\/} spinors of the same type
are proportional to each other; this cannot be 
true in general, because they are independent.

\end{itemize}

Applying these rules to the four-point double-box does indeed give the six
allowed solutions of the two types shown in \fig{heptacutFigure}.
The remaining solutions are
related to the depicted ones by flip symmetries.  
The complete set of solutions to the cut constraints is solely determined by 
the topology of a given diagram.  Each solution determines a pattern
of\break
`\PlusVertex' and `\MinusVertex' vertices in the diagram.  
For each
solution to the cut constraints, one of the two types of three-point
amplitudes vanishes at each vertex.  This
pattern (along with the topology)
will in general restrict the helicity assignments along internal lines,
and may also restrict the particle
types allowed in different internal lines.

The strongest constraint that can arise in a kinematic
solution is the restriction to a single allowed helicity configuration
for the internal lines. We will refer to this configuration as a
``singlet''.  In this case only gluons can propagate inside the
diagram, as in \fig{singlet}(a).   Fermions or scalars are not allowed
because the only potentially non-vanishing vertices are of the 
wrong type and vanish for the given solution. 
The second-strongest constraint
allows two helicity configurations. In such
configurations the particle content is purely gluonic except for one
loop in which any particle type can propagate, as shown in
\fig{singlet}(b). (A fermionic loop always allows two helicity
assignments, corresponding to interchanging fermion and antifermion,
and the same is true for complex scalar loops.)

Solving for the spinors in any diagram is then straightforward.
Consider the singlet case, diagram (a)
in \fig{singlet}.   The on-shell conditions together with
momentum conservation at each vertex give a set of equations
that must be satisfied,
\begin{eqnarray}
&&\lambda_{k_1}  \propto \lambda_{l_3}   \propto \lambda_{l_1} \,,
\hskip 2 cm \lambda_{k_1} \tlambda_{k_1} = \lambda_{l_3}
\tlambda_{l_3} -
   \lambda_{l_1}\tlambda_{l_1} \,,
\nn \\
&&\tlambda_{k_2} \propto \tlambda_{l_1} \propto \tlambda_{l_2}\,,
\hskip 2 cm \lambda_{k_2} \tlambda_{k_2} = \lambda_{l_1}
\tlambda_{l_1} -
   \lambda_{l_2}\tlambda_{l_2} \,,
\nn \\
&&\tlambda_{k_3} \propto \tlambda_{l_5} \propto \tlambda_{l_7}\,,
\hskip 2 cm \lambda_{k_3} \tlambda_{k_3} =
\lambda_{l_5}\tlambda_{l_5}- \lambda_{l_7}
\tlambda_{l_7} \,, \\\
&&\lambda_{k_4} \propto \lambda_{l_7} \propto \lambda_{l_6}\,,
\hskip 2 cm \lambda_{k_4} \tlambda_{k_4}
=\lambda_{l_7}\tlambda_{l_7}- \lambda_{l_6}
\tlambda_{l_6} \,,\nn \\
&&\tlambda_{l_4} \propto \tlambda_{l_6} \propto \tlambda_{l_3}\,,
\hskip 2.2 cm \lambda_{l_4} \tlambda_{l_4} =
\lambda_{l_6}\tlambda_{l_6}-\lambda_{l_3}
\tlambda_{l_3} \,,\nn \\
&&\lambda_{l_4} \propto \lambda_{l_5} \propto \lambda_{l_2} \,,
\hskip 2.2 cm \lambda_{l_4} \tlambda_{l_4} = \lambda_{l_5}
\tlambda_{l_5} -
   \lambda_{l_2}\tlambda_{l_2} \nn
\end{eqnarray}
The solution to these equations is,
\begin{equation}
\begin{array}{lcl}
\lambda_{l_1}=\lambda_{k_1}\,, & \phantom{space} &
\widetilde{\lambda}_{l_1}=\xi \widetilde{\lambda}_{k_2} \,,\\
\lambda_{l_2}=\xi \lambda_{k_1}- \lambda_{k_2}\,, &\phantom{space} &
\widetilde{\lambda}_{l_2}=\widetilde{\lambda}_{k_2}\,, \\
\lambda_{l_3}=\lambda_{k_1} & \phantom{space} &
{\tlambda}_{l_3}=\xi {\tlambda}_{k_2}+{\tlambda}_{k_1} \\
\lambda_{l_4}=\lambda_{l_2} & \phantom{space} &
\widetilde{\lambda}_{l_4}=\widetilde{\lambda}_{l_3}
 \frac{\vphantom{\tilde A}\spb2.3}{\spb3.{l_3}}  \\
\lambda_{l_5}=\lambda_{l_2} &\phantom{space} &
\widetilde{\lambda}_{l_5}=\widetilde{\lambda}_{k_2}+\widetilde{\lambda}_{l_4}\\
\lambda_{l_6}=\lambda_{k_1}+ \lambda_{l_2}
\frac{\spb2.3}{\spb3.{l_3}} &\phantom{space} &
\widetilde{\lambda}_{l_6}=\widetilde{\lambda}_{l_3}\\
\lambda_{l_7}=\lambda_{k_4} &\phantom{space} &
\widetilde{\lambda}_{l_7}=\widetilde{\lambda}_{k_3}
\frac{\spa3.{l_2}}{\spa{l_2}.4}\\
\end{array} \label{solution}
\end{equation}
where $\xi$ is an arbitrary parameter, corresponding to the
remaining degree of freedom in the integration not frozen by the
hepta-cut. Since a bispinor $p^{a \dot{a}}=\lambda^a
\widetilde{\lambda}^{\dot{a}}$ is invariant under a rescaling 
of the spinors, $(\lambda^a
,{\tlambda}^{\dot{a}}) \rightarrow (\beta \lambda^a\ ,\beta^{-1}
{\tlambda}^{\dot{a}})$, the above solution can be written in many
other forms. In addition, there is a choice as to where to include
the remaining degree of freedom.  While individual
three-point amplitudes $\VT{\pm}$  are not invariant under this
transformation, the product of amplitudes forming the cut is invariant.

\subsection{Solving for integral coefficients using maximal cuts}
\label{CoefficientSolutionSubsection}

We now consider how to solve for the coefficient of an integral
using the maximal cuts.  At two loops, there is only a single conformal
integral, the double box.  It can appear, of course, in both $s$-
and $t$-channel configurations, but the hepta-cut shown in
\figs{heptacutFigure}{singlet} selects only the $s$-channel double box.
Our candidate expression for the amplitude is then,
\begin{equation}
A^{(2)}(1,2,3,4) = c \, A^{\tree}(1,2,3,4) I^\twoloop(s,t)\,,
\end{equation}
where $I^\twoloop(s,t)$ is the pseudo-conformal two-loop double-box
integral in \eqn{2box} and $c$ is a coefficient that we need to
solve for.  This integral contains a factor of $s^2 t$ in the
numerator, which as we shall see is necessary for satisfying the cut
conditions.

Imposing the sevenfold cut condition, we obtain,
\begin{eqnarray}
&& c s^2 t A^\tree(1,2,3,4) \int d^4 l_1  d^4 l_7
        \prod_{i=1}^7 \delta(l_i^2) \nn \\
&& \hskip 2 cm \null
 = i  \int d^4 l_1  d^4 l_7  \prod_{i=1}^7 \delta(l_i^2)
\sum_{h} (A^{\tag1}_{(1)} A^{\tag2}_{(2)}A^{\tag3}_{(3)}
A^{\tag4}_{(4)}A^{\tag5}_{(5)}A^{\tag6}_{(6)})_h \,,
\label{heptacut}
\end{eqnarray}
where $A^{\tag{i}}_{(i)}$ is the three-point amplitude corresponding
to one of the six three vertices and the sum over helicities $h$ runs
over all possible helicity and particle configurations.  (We have
taken the loop integrals to be four-dimensional for the purposes of
our discussion here, but in any explicit evaluation of the integrals
they should be continued to $D$ dimensions to regulate the infrared
singularities.)

As discussed above, the delta-function constraints are solved
by a discrete set of solutions, so we obtain
\def\sol#1{{{\rm sol}_{#1}}}
\begin{eqnarray}
&& \hskip -.4 cm 
  c \, s^2 t A_4^\tree(1,2,3,4) \int d^4 l_1  d^4 l_7
      \int \d \xi \, \sum_{j=1}^6  J_j\, \delta^4(l_1 - l_1^\sol{j})
         \, \delta^4(l_7 - l_7^\sol{j}) \label{heptacutsol} \\
&& \hskip .4 cm 
 = i  \int d^4 l_1  d^4 l_7  \int \d \xi \, 
  \sum_{j=1}^6 J_j \, \delta^4(l_1 - l_1^\sol{j})
         \, \delta^4(l_7 - l_7^\sol{j}) 
\sum_{h \in H_j} (A^{\tag1}_{(1)} A^{\tag2}_{(2)}A^{\tag3}_{(3)}
A^{\tag4}_{(4)}A^{\tag5}_{(5)}A^{\tag6}_{(6)})_{j,h} \,, \nn
\end{eqnarray}
where $j$ runs over the different kinematic solutions, $l_1^\sol{j}$
and $l_7^\sol{j}$ are the values of the independent loop momenta 
expressed in terms of the external momenta, and the remaining degree of
freedom is $\xi$.  For each discrete solution $j$, only a subset of
helicity and particle configurations denoted by $H_j$ gives a
non-vanishing contribution. The Jacobian from the change of variables
is $J_j$. 

Buchbinder and Cachazo~\cite{BuchbinderCachazo}
noted that the integrand is constant after
imposing the seven cut conditions arising directly from cutting propagators,
without need to impose the eighth cut condition.
Another curiosity they noted
is that all six discrete kinematic solutions for
the hepta-cut give the same answer for the amplitude. 
This was true for kinematic
solutions that permitted only gluons in the two loops as well as for those
which also permitted fermions and scalars.  This simplicity
is related to the absence of terms which integrate to zero 
upon performing the loop integral.\footnote{This property is special 
to four-point amplitudes
and is already violated at one loop for five-point amplitudes.}

We will exploit this observation, and assume its generalization to
higher loops.  It allows us to match integrands, and indeed to pick
individual solutions to match the left-hand and right-hand side of
\eqn{heptacutsol}, determining the overall coefficient.
That is, we assume that there is a single overall coefficient $c$ to
solve for in front of each integral, instead of a different contribution
for each solution.
This also avoids any need for integral
reductions or analysis of the integrals, and translates integrands
into algebraic coefficients of integrals.  Our 
knowledge of an integral basis --- given by the
pseudo-conformal integrals --- is not essential but greatly simplifies the 
extraction of these coefficients.  This equality of 
contributions from different solutions
is likely special to
MSYM at four points
or perhaps to conformal supersymmetric gauge theories more generally.
In general, there is no reason to expect solutions which allow different
particle types to circulate to yield equal answers.

The assumption can be checked directly, of course, by comparing 
different solutions.  While we have not checked it exhaustively,
it does pass the large number of such comparisons that we have carried out.  
The use
of the assumption and the maximal-cut procedure described here also leads
to a determination of coefficients at three and four loops in agreement
with known answers~\cite{BDS,BCDKS}.
Furthermore, 
a violation would likely lead to inconsistent determinations of 
integral coefficients at five loops; we find no such inconsistency.
We can also rely on cross checks from non-maximal cuts.

(The reader may be puzzled by the appearance of complex solutions in
what was originally an integral over real loop momenta.  This is not
special to the amplitudes under consideration here.  In extracting the
cut by replacing propagators with delta functions, one must sum over
complex solutions as well as real ones~\cite{BCFUnitarity}.  This was
necessary in other circumstances such as evaluating the connected
prescription for tree-level gauge-theory amplitudes~\cite{RSV} in
twistor string theory~\cite{WittenTopologicalString}.  It can also
be understood by reinterpreting~\cite{Vergu} the original integral as
a fourfold contour integral in each component of the loop momentum,
and replacing the propagators by products of an expression of the form $[2\pi
i(l_i^\mu-l_i^{\mu, {\rm sol}_j})]^{-1}$ times Jacobians; 
Cauchy's theorem makes it
act like a delta function, but allowing complex solutions.
The details are not important to us here because we are only determining
coefficients and not evaluating any integrals.)

Choosing one of the kinematic solutions then gives,
\begin{equation}
c \, s^2 t A_4^\tree(1,2,3,4) = i \sum_h(A^{\tag1}_{(1)}A^{\tag2}_{(2)}
A^{\tag3}_{(3)} A^{\tag4}_{(4)}A^{\tag5}_{(5)}
A^{\tag6}_{(6)})_h \,,
\label{ustatement}
\end{equation}
where $h$ runs over the helicity configurations and particle content
with non-vanishing contributions for the given solution. 

Obviously, it is advantageous to choose the simplest solution, where
the kinematics restricts us to the fewest possible particle types
circulating in the loop.  The best choice is a singlet where only
gluons contribute. Using the kinematic solution~(\ref{solution}),
corresponding to the singlet solution in \fig{singlet}(a), we have,
\begin{equation}
 c\, s^2 t A_4^\tree(1^-,2^+,3^+,4^-)  = i A^{\tag1}_{(1)}
A^{\tag2}_{(2)}A^{\tag3}_{(3)}A^{\tag4}_{(4)}
A^{\tag5}_{(5)}A^{\tag6}_{(6)}\,,
\label{singletstatement}
\end{equation}
with no sum over intermediate helicities.  In the singlet case there
is only one term in the helicity sum and all six three-point
amplitudes are purely gluonic.  Here, $A^{\tag{j}}_{(j)}$ represents
one of the three-gluon tree amplitudes in
\eqns{ThreeVertexMinus}{ThreeVertexPlus}, with the plus and minus
labels on these amplitudes matching the labels of the vertices in the
figure.  We have confirmed that \eqn{singletstatement} holds for any
value of the arbitrary parameter $\xi$ (other than $\xi = 0$, where 
the right-hand-side of \eqn{singletstatement}
is ill defined) in the kinematic solution
(\ref{solution}).  This equation then determines $c=+1$, so that the
pseudo-conformal double box integral in \fig{dual2loop} appears in the
two-loop amplitude with a coefficient of
$+A_4^\tree(1^-,2^+,3^+,4^-)$, in agreement with known
results~\cite{BRY,BDDPR}.


\subsection{Generalized unitarity with real momenta at two loops}
\label{GeneralizedSubsection}

\begin{figure}
\centerline{\epsfxsize 2.8 truein \epsfbox{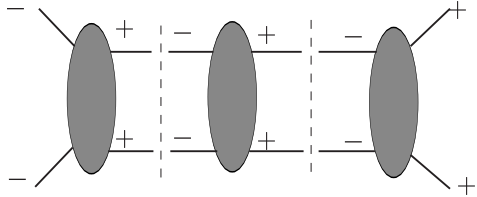}}
\caption{A singlet iterated two-particle cut of the two-loop four-point
amplitude.}
 \label{IteratedTwoLoopCutsFigure}
\end{figure}

The kinematics in maximal cuts is highly constrained.
It is therefore useful to have a way of checking results using 
less-restricted kinematics.  Let us begin by considering
generalized (non-maximal) cuts which are well defined for real
momenta in four dimensions.

As an example of a four-dimensional generalized cut, consider the
two-loop iterated two-particle cuts shown in
\fig{IteratedTwoLoopCutsFigure}.  This helicity configuration has the
property that it is a singlet under supersymmetry
transformations~\cite{BDDPR}, the only contributions coming from gluon
internal states.  Hence we call it the ``singlet'' contribution.  The
remaining contributions, containing all other helicity and
particle-type assignments, we will collectively call the
``non-singlet'' contributions.  These latter contributions transform
into each other under supersymmetry.  (The action of supersymmetry on
the $\NeqFour$ amplitudes is described in appendix E of
ref.~\cite{BDDPR}.)

\begin{figure}
\centerline{\epsfxsize 2.5 truein \epsfbox{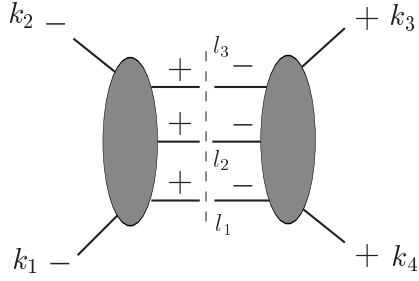}}
\caption{A singlet contribution to the two-loop three-particle
cut. Only gluons enter in the loops.} \label{TwoLoopThreeCutFigure}
\end{figure}

The singlet configuration is especially simple to evaluate because it
involves only a single particle type, similar to the singlet
configurations of the maximal cuts. At higher loops, the number of
different particle and helicity configurations grows rapidly.  If
possible, it would be simpler to use only singlet configurations to
confirm our ansatz for the amplitude, just as with maximal cuts.
Unlike maximal cuts, however, the generalized cuts considered here
do not enforce a particular choice of internal-line
helicities or particle types.
Nonetheless, in special cases, the singlet can be used to
determine the coefficient of integrals in the amplitude.

For example, in the iterated two-particle cuts at two loops shown
in \fig{IteratedTwoLoopCutsFigure}, the singlet contribution gives exactly
the coefficient of the double box integral~\cite{BRY}.  With this
external helicity configuration, this cut has no non-singlet
contribution.  The non-singlet contribution appears in the other
channel and gives the identical result.  In the three-particle cut,
however, the singlet and non-singlet contributions appear in the same
cut and are not identical.  This cut does nonetheless have simple
properties that we can exploit.  The singlet contribution, depicted in
\fig{TwoLoopThreeCutFigure}, has been previously evaluated in
refs.~\cite{BRY,BDDPR}, with the result,
\begin{eqnarray}
C^{\rm singlet} &=& A_5^{\rm tree} (1^-, 2^-, l_3^+, l_2^+, l_1^+)
\times A_5^{\rm tree} (3^+, 4^+, -l_1^-, -l_2^-, -l_3^-) \nn \\
& = & - \spa1.2^2 \spb3.4^2
 {\tr_+[1 l_1 43 l_3 2] \over
 (l_1 + l_2)^2 (l_2 + l_3)^2 (l_3 - k_3)^2
(l_1 - k_4)^2 (l_3 + k_2)^2 (l_1 + k_1)^2} \,, \hskip 1 cm
\label{TwoLoopSingletCut}
\end{eqnarray}
where $\tr_\pm [1 l_1 \cdots] = \tr [(1\pm\gamma_5) \ksl_1 \lsl_1 \cdots]/2$.
The tree amplitudes that appear are,
\begin{eqnarray}
A_5^{\rm tree} (1^-, 2^-, l_3^+, l_2^+, l_1^+) & = &
i {\spa1.2^4 \over \spa1.2 \spa2.{l_3} \spa{l_3}.{l_2}
      \spa{l_2}.{l_1} \spa{l_1}.1} \,, \nn \\
A_5^{\rm tree} (3^+, 4^+, -l_1^-, -l_2^-, -l_3^-) & = &
- i {\spb3.4^4 \over \spb3.4 \spb4.{(-\! l_1)} \spb{(-l_1)}.{(-\!l_2)}
      \spb{(-l_2)}.{(-\!l_3)} \spb{(-l_3)}.3}\,. \hskip 1.5 cm
\end{eqnarray}
The non-singlet contribution arising from the contribution of all
other helicity and particle configurations crossing the cut is a bit
more complicated to evaluate, and is equal to~\cite{BRY,BDDPR},
\begin{eqnarray}
C^{\rm non\hbox{-}singlet}
&=& -\spa1.2^2 \spb3.4^2
 {\tr_-[1 l_1 43 l_3 2] \over
 (l_1 + l_2)^2 (l_2 + l_3)^2 (l_3 - k_3)^2
(l_1 - k_4)^2 (l_3 + k_2)^2 (l_1 + k_1)^2} \,.  \hskip 1 cm
\end{eqnarray}
The $\gamma_5$ terms in the singlet and non-singlet appear
with opposite signs.  In the sum over singlet and non-singlet
contributions to the cut, the $\gamma_5$ terms therefore cancel
algebraically at the level of the integrand.  (Alternatively,
the difference between the singlet and non-singlet contributions
integrates to zero.)  From a practical standpoint, it is easier to
compare cuts with target ans\"{a}tze prior to integration and to use
only the singlet, so the key observation is that we may use is that
the non-$\gamma_5$ term of the singlet is exactly half the total
contribution to the cut.

We need to extend these observations to higher loops in order for them 
to be useful. We have confirmed that the same properties hold
at three and four loops for any combination of two- and three-particle
cuts composed only of four- and five-point tree amplitudes. (If six-
or higher-point tree amplitudes are present in the cuts, non-MHV
configurations with three or more negative and three or more positive
helicities enter into the computation, which renders the structure of
supersymmetric cancellations more elaborate and prevents us from using
the singlet contribution alone to evaluate the cut.)  We will assume that
this observation continues to hold true at five loops.  With this
assumption we will be able to check (in \sect{ConfirmingCutsSection})
the coefficients of a variety of pseudo-conformal integrals, using only
singlet cuts. This is a strong consistency check, because it relies not
only on the coefficient under examination being correct, but also on
the assumption remaining valid.
(It seems extremely implausible that a breakdown of the assumption
at five loops could be compensated by an incorrect coefficient.)

To do better we need to sum over all states crossing the cuts.
Moreover, a proper treatment of the cuts requires that 
the cuts be evaluated using $D$-dimensional states and 
momenta~\cite{DDimUnitarity,OneLoopReview,DimShift}.  This ensures
that no contributions have been dropped, as can happen when
four-dimensional momenta are used.  At one loop, the improved power
counting of supersymmetric theories allows one to prove a theorem that
unitarity cuts with four-dimensional momenta are sufficient to
determine dimensionally-regulated supersymmetric amplitudes (that is,
``near'' four dimensions) completely~\cite{NeqFourOneLoop,Fusing}.  (The
regulator must of course maintain manifest supersymmetry; as mentioned
earlier, we use the four-dimensional helicity scheme (FDH) to do so.
In this scheme, the helicity algebra is always four dimensional, but the
momenta are continued to $D=4-2\e$ dimensions.)  Unfortunately, no
such theorem is as yet known beyond one loop.  A subtlety in deriving
such a theorem arises from infrared singularities: the singularities
in one loop can effectively ``probe'' the $\Ord(\ep)$ contributions
from another loop, the product giving a surviving contribution even as
$\ep \rightarrow 0$.  When computing with $D$-dimensional momenta, one
can no longer use the spinor helicity
representation~\cite{SpinorHelicity}, which makes expressions for tree
amplitudes used in the cuts more complicated.  A good way to
ameliorate this additional complexity is to consider instead $\NeqOne$
in ten dimensions super-Yang-Mills dimensionally reduced to
$D=4-2\eps$ dimensions.  The remaining states are completely
equivalent to those of MSYM in the FDH scheme~\cite{FDH}, except that
the bookkeeping of contributions is much simpler.

At two loops all cuts of MSYM amplitudes were evaluated in $D$
dimensions~\cite{TwoLoopGluons}, providing a complete proof of the
planar and non-planar expressions for the MSYM amplitudes first
obtained in ref.~\cite{BRY} using four-dimensional momenta.  At three
loops, we have also re-evaluated the planar amplitude using
$D$-dimensional cuts.  The four-loop planar amplitude has been evaluated
using $D$-dimensional cuts, assuming that the full result (as an abstract
tensor in polarization vectors and momenta) is proportional to the
tree amplitude, and making the 
reasonable assumption that no contributions can have a triangle
subintegral.  (Only the terms involving polarization vectors dotted into
each other, after tensor reductions, were evaluated explicitly.
Also, one can rule out all bubble and some triangle subintegrals 
using supersymmetry along with generalized unitarity.)  
In \sect{ConfirmingCutsSection}, we will make use
of these results to provide non-trivial
evidence in favor of the various assumptions we have used to obtain
our ansatz for the five-loop four-point planar amplitude.  A complete
proof would require additional $D$-dimensional cuts be evaluated in
order to confirm the coefficient of every potential integral that
might appear, including non-conformal ones.


\section{Maximal Cut Technique for Determining Integral Coefficients}
\label{MaximalCutSection}

\subsection{Overview of maximal cut method}
\label{OverviewSubsection}

In this section, we further develop the maximal-cut technique for higher
loops using the observations of the previous section.  This allows us
to extend the two-loop maximal cuts of Buchbinder and
Cachazo~\cite{BuchbinderCachazo} to higher-loop orders.  Unlike
Buchbinder and Cachazo, we do not require the loop integration be
frozen by the cut conditions. That is, we do not require the number of
cut conditions to match the number of loop integrations.  As
discussed in the previous section, instead, we
perform all evaluations of the cuts at the level of the integrand,
prior to performing any loop integrations.

Moreover, we do not solve for all possible kinematic configurations
satisfying the cuts.  We instead focus on those solutions which allow
the simplest determinations of the coefficients of the
pseudo-conformal integrals as they appear in the amplitude.  As
discussed in \sect{MaximalCutsSubsection}, the simplest kinematic
solutions are the singlets, to which only gluons contribute.  At five
loops it turns out that only four pseudo-conformal integrals do not
have singlet solutions.  However, even in these cases one can choose
kinematics which forces the fermion and scalar contributions into
specific loops, again greatly simplifying the determination of the
coefficients.  In order to speed up the process of extracting the
coefficients we solve the constraint equations numerically, although
in some cases we find it useful to solve the constraints analytically.

At two-, three- and four-loops, where the complete results for the
four-point planar amplitudes are known~\cite{BRY,BDS,BCDKS}, we have
confirmed that singlet maximal cuts correctly determine the
coefficients of all integrals appearing in the amplitudes.  This
suggests that the same will be true at five loops.  Again, if this
were not true it would reveal itself as an inconsistency in the
results.  In particular, we would find that different kinematic
solutions of the cut conditions would lead to inconsistent determinations
of integral coefficients.  We would also find inconsistencies 
with cuts with less restrictive kinematics.

A drawback of our maximal-cut method is its reliance on
four-dimensional spinor helicity, which as mentioned in
\sect{GeneralizedSubsection} might drop contributions.  Nevertheless, it
does provide a relatively simple and systematic means to obtain an
ansatz for the coefficients of integrals that appear in the amplitude.

To evaluate the coefficient of any of the five-loop integrals with
only three-point vertices shown in \fig{cubicsFigure}, we cut all 16
propagators. Similarly, the coefficient of integrals with quartic
vertices in \fig{quarticsFigure}, can be obtained by cutting all of
the propagators present.  This is of course fewer than the number of
propagators present in diagrams with only cubic vertices, so some of the
integrals with only cubic vertices can contribute to the cut.  We
must therefore subtract out all such contributions, to obtain the
coefficient of a particular integral containing four-point vertices.  For
some some solutions, the kinematics does not allow the coefficients of
integrals with quartic vertices to be determined.  For
example, if the kinematic constraints due to three-point vertices
force the spinors of two nearest neighboring legs of the four-point
subamplitude to be proportional, $\lambda_i\propto\lambda_j$, the sum of
momenta of these legs will be on-shell, $(k_i + k_j)^2 = 0$. This can
place an internal propagator of the four-point subamplitude on-shell, where 
it diverges.  This effectively selects out only those terms with this
propagator present and loses the four-point contact contribution.  In
such cases, we determine the integral coefficient by using
a different solution to the cut conditions.

\subsection{Evaluating the five-loop integral coefficients}

A maximal cut for determining the coefficient of integrals with
a given set of propagators is of the form,
\begin{equation}
C^\mathrm{\fiveloop} \Bigr|_\mathrm{maximal}  = i^c \sum_h
\biggl(\prod_{k=1}^{m} A^{\tree}_{(k)}\biggr)_h, \label{genunit}
\end{equation}
where $h$ signifies the different helicity configurations and particle
types that can contribute and $m$ is the number of tree amplitudes
appearing in the cut. In this equation the cut momenta $l_1, l_2,
\ldots, l_c$ are all on shell.  Euler's formula relates the number of
tree amplitudes that appear to the number of cut lines; at five loops
$m = c -4$. As discussed in \sect{MaximalCutsSubsection}, 
a given kinematic solution to
the cut conditions will allow only a subset of helicity configurations
to contribute.

\begin{figure}[t]
\centerline{\epsfxsize 5.8 truein \epsfbox{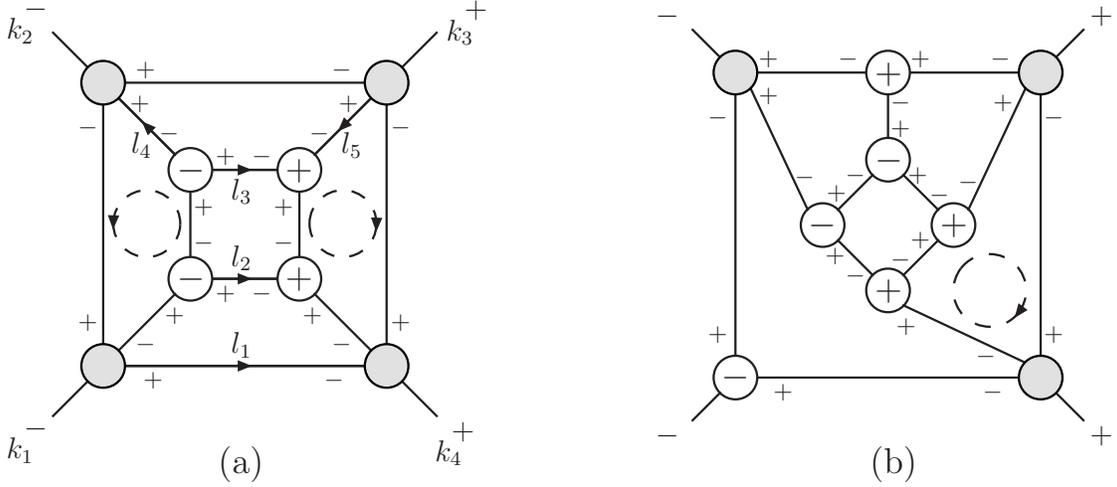}}
\caption{(a) The simplest cut determining the coefficient of
 $I_{37}$.
(b) A particularly good choice of cut for determining
the coefficient of $I_{39}$.
All lines are cut and carry on-shell momenta.
In all loops except those indicated by a
dashed circle, only gluons propagate; the dashed circle indicates
that all particle types can circulate. Other allowed helicity
configurations are obtained by flipping all helicities in these
loops. Grey blobs represent four-point amplitudes, which hide 
possible propagators inside.}
 \label{nonsingletFigure}
\end{figure}

The simplest situation is when an integral has only cubic vertices,
as is true for the integrals 
of~\fig{cubicsFigure}.  In this case one can always choose
singlet kinematic solutions to the on-shell conditions so that
only gluons propagate in each loop. After cutting all the propagators,
one obtains the numerator $N$ subject to the cut conditions,
\begin{equation}
N= \biggl(\prod_{j=1}^{12}
A^{\tag{j}}_{(j)}\biggr)_\mathrm{singlet},
\phantom{extraspace} (l^2_1, l^2_2, \ldots, l^2_{16} = 0)
\label{cubicsinglet}
\end{equation}
where the tree amplitudes $A^{\tag{j}}_{(j)}$ are purely gluonic. In
some cases, such as cuts isolating integrals $I_1$ or $I_2$, only a
single term appears, but in others a sum of terms appear.
For example, there are five
contributions to the cut with propagator configuration of $I_{21}$
and $I_{22}$; the integral $I_{21}$ appears four times in the cut as
well as in the amplitude, but with the numerator factor permuted,
while $I_{22}$ appears one time.


\subsection{Examples of evaluations of integral coefficients}

As a first example of the determination of the coefficient of an
integral, consider the maximal cut of the ``ladder'' integral $I_1$.
Here $ N= i st^5 A^\tree_4(1,2,3,4)$, which is independent of the loop
momenta. This coefficient was determined long ago, from iterated
two-particle cuts~\cite{BRY}.  To confirm this result using maximal
cuts we solve the on-shell constraints $l_i^2$ of all sixteen cut
propagators. For the external helicities $(1^-,2^-,3^+,4^+)$ there are
166 distinct kinematic solutions of which 62 correspond to singlet
configurations.  For $(1^-,2^+,3^-,4^+)$ there are 270
solutions of which 27 are singlets, and for $(1^-,2^+,3^+,4^-)$ the
numbers are similar: 270 solutions, with 28 singlets. We have
confirmed that all these singlet solutions individually agree with the known
result, providing a non-trivial check. 
This determines that the pseudo-conformal integral $I_1$ enters 
with an overall factor of $-1/32$ in the normalized amplitude $M_4^{(5)}$
given in \eqn{FiveLoopAnsatz}, after accounting for the normalization
conventions of the integrals in \eqn{IntegralNormalization}.
(The $1/32$ prefactor in \eqn{FiveLoopAnsatz} does not appear in the
product of tree amplitudes making up the cut, but appears in the
$L$-loop amplitude, due to our convention of including a factor of
$2^L$ in \eqn{LeadingColorDecomposition}.)  In the integrals of
\fig{cubicsFigure} and \fig{quarticsFigure} we have not included the
overall factor of $-1/32$, but leave it as an explicit overall factor
in \eqn{FiveLoopAnsatz}.  In the remaining part of this section, we
will refer only to signs relative to $I_1$.

As a second example, consider $I_{17}$ and $I_{18}$ in
\fig{cubicsFigure} as well as $I_{46}$ in \fig{nonSTIntegralsFigure}.
They have the same propagators and hence can contribute to same
maximal cut.  For external helicities $(1^-,2^-,3^+,4^+)$ there are
335 kinematic solutions of which 62 are singlets. The helicity
configuration $(1^-,2^+,3^-,4^+)$ has 339 solutions and 48 singlets
whereas $(1^-,2^+,3^+,4^-)$ has 304 solutions and 102 singlets. Again
all singlet solutions individually give results consistent with the
coefficient of $I_{46}$ vanishing and both $I_{17}$ and $I_{18}$
entering the amplitude with a numerical coefficient of $+1$ relative
to $I_1$.

As a third example, consider a maximal cut of integral $I_{21}$.
Together with integrals having the same propagators, $I_{22}$,
$I_{35}$ and $I_{49}$, there are nine potential terms to the numerator
$N$ after symmetrization.  For this cut the helicity configuration
$(1^-,2^-,3^+,4^+)$ has 376 solutions of which 98 are singlets and
$(1^-,2^+,3^-,4^+)$ has 384 solutions of which 58 are singlets. Note
that helicity $(1^-,2^+,3^+,4^-)$ is related to $(1^-,2^-,3^+,4^+)$,
by symmetry. It turns out that the singlets can never be
made consistent with numerators of type $I_{49}$, hence its
coefficient must vanish.  But unfortunately, the singlet cuts do not
uniquely fix the coefficients of the remaining integrals.  For
example, if we were to assume relative numerical coefficients of $\pm
1$, there are exactly two possibilities, one involves five terms given
by symmetrizations of numerators of type $I_{21}$ and $I_{22}$ and the
other involves two terms of type $I_{35}$.  To resolve this situation
we must instead consider cuts with fewer cut conditions imposed, to
reduce the degeneracy of the kinematics.

Maximal cuts of diagrams involving non-cubic vertices are only a bit
more complicated. Luckily almost all cuts needed to determine the
coefficients of the integrals have singlets in their solution set.
Only $I_{37}$, $I_{39}$, $I_{55}$ and $I_{59}$ in
\figs{STIntegralsFigure}{nonSTIntegralsFigure} do not have singlet
solutions.  For these cases we must use non-singlet cuts,
such as those in \fig{nonsingletFigure}.

As an example of a singlet solution with a quartic vertex, consider a
maximal cut of $I_{32}$. An expression that correctly matches the
singlet maximal cut is \hbox{$C_4^\fiveloop = i A^\tree_4(1,2,3,4)(s^2t^2+
\ldots)$}, where ``$\ldots$'' stands for 14 rational terms obtained
from integrals $I_{6}$, $I_{11}$, $I_{12}$, $I_{21}$, $I_{22}$ and
$I_{31}$, which also contribute to this cut. Since the coefficients of
these integrals can be determined from other cuts, we simply subtract their
contributions allowing us to determine the coefficient of $I_{32}$ to be 
$+1$.  There are now fewer cubic vertices in the cut and consequently
the number of kinematic solutions also drops: The three inequivalent
external helicity arrangements each have 18 solutions that are not
degenerate in the two four-point blobs. However, they differ in their
singlet content: Helicities $(1^-,2^-,3^+,4^+)$ have four singlets,
$(1^-,2^+,3^-,4^+)$ have no singlets, and $(1^-,2^+,3^+,4^-)$ have
exactly one singlet.  (When solving for the kinematics, we do not include
solutions which do not allow us to determine $I_{32}$, because the
four-point contact terms are lost, as mentioned in~\sect{OverviewSubsection}.)

As mentioned above, integrals $I_{37}$, $I_{39}$, $I_{55}$ and
$I_{59}$ have no singlet solutions in their maximal cuts. For $I_{37}$
an useful choice is to force scalars and fermions to circulate in only
two independent non-overlapping loops; there is only a single
kinematic solution with this property, which helicity configuration
shown in~\fig{nonsingletFigure}(a).  $I_{39}$, $I_{55}$ and $I_{59}$
all have simpler solutions with only one loop that carries fermions
and scalars. For $I_{39}$, for example, three kinematic solutions exist
with this property, one of which is displayed in
\fig{nonsingletFigure}(b).

A cut where a fermion or scalar can
circulate in only one of the loops takes the form,
\begin{equation}
C_4^\mathrm{5-loop} = i^c \sum_{h\in \{+,-\}} \biggl\{
\biggl(\prod_{j=1}^{c-4} A^{\tag{j}}_{(j)}\biggr)_\mathrm{\! gluon} 
-4 \biggl(\prod_{j=1}^{c-4} A^{\tag{j}}_{(j)}\biggr)_\mathrm{\!fermion} 
+ 3 \biggl(\prod_{j=1}^{c-4}
A^{\tag{j}}_{(j)}\biggr)_\mathrm{\! scalar} \biggr\} \label{doublet}
\label{NonSingletLoop}
\end{equation}
where $h$ is the helicity of the particle in the unique loop with
fermions and scalars. (Helicity is conserved along this loop, which
given our all-outgoing convention means that it flips going from one
vertex to the next.)  For the complex scalars, the two helicities
correspond to particle and antiparticle.  For the cut in
\fig{nonsingletFigure}(b) only four vertices involve particles other
than gluons, hence five $A^{\tag{j}}$'s can be pulled out of the sum
as a common factor.  The required four-point tree amplitudes for
different particles follow from the supersymmetry Ward
identities~\cite{SWI}, which are described, for example, in Appendix E
of ref.~\cite{BDDPR}.  From this cut we find that $I_{39}$ does not
contribute to the amplitude. Likewise the maximal cuts of $I_{55}$ and
$I_{59}$ shows that coefficients of these integrals vanish.

For the cut in \fig{nonsingletFigure}(a) one can
arrange the kinematics so that the two loops that can carry fermions or
scalars do not intersect, simplifying their evaluation.  The structure 
is similar to \eqn{NonSingletLoop}, except that there are two independent
sums over fermions, scalars and gluons.
We will not present it explicitly, but instead just give
the kinematic solution needed for this cut,
\begin{equation}
\begin{array}{rlcrl}
\displaystyle  l_1&\displaystyle=p=\lambda_p \widetilde{\lambda}_p\,,
&\hskip 10mm&l_2&\displaystyle=q=\lambda_q \widetilde{\lambda}_q\,,\\
\displaystyle  l_3&\displaystyle= 
   -\lambda_q \widetilde{\lambda}_q {(p+q+k_1+k_2)^2 \over
\sand{q}.{p+k_1+k_2}.{q}}\,,
&\hskip 10mm&\displaystyle l_4&\displaystyle= 
   -\lambda_q \widetilde{\lambda}_r \frac{(p+q+l_3+k_1)^2}{
\sand{q}.{p+k_1}.{r}}\,, \\
\displaystyle  l_5&\displaystyle= 
   -\lambda_v \widetilde{\lambda}_q \frac{(p+q+l_3-k_4)^2}{
\sand{v}.{p-k_4}.{q}}\,.
\end{array} 
\label{nonsingsolution}
\end{equation}
Here $p$ and $q$ are arbitrary null vectors in four dimensions and
$\widetilde{\lambda}_r$ and $\lambda_v$ are spinors corresponding
to arbitrary null vectors $r,v$. The
remaining seven loop momenta can be obtained by
momentum conservation. 

This cut is the least discriminating one needed for fixing the
coefficients of the integrals in the five-loop amplitude, and hence it
contains the most terms. There are 79 terms of the right conformal
weight that are candidates for the left-hand-side of~(\ref{genunit}),
but of these only 28 terms contribute to the amplitude. These terms
are obtained from integrals $I_5$, $I_{16}$, $I_{20}$, $I_{21}$,
$I_{22}$ and $I_{27}$; the coefficient of $I_{37}$ must thus vanish.
Interestingly, this is the only maximal cut where integrals
($I_{21}$ and $I_{22}$) enter twice compared to their appearance in
the amplitude.

In some cases the maximal cuts cannot distinguish between different
integrals, due to the degenerate nature of the kinematics.  As an
example, consider the maximal cut of $I_{21}$, $I_{22}$ and $I_{35}$
described above, which has two possible numerator combinations
satisfying the cut conditions.  On the maximal cut of these diagrams
we find,
\begin{equation}
(I_{21}+I_{22}
 -I_{35})\Bigl|_\mathrm{cut}=0 \,.
\label{PotentialAmbiguity2}
\end{equation}
The combination of $I_{21}$ and $I_{22}$ make one possible numerator
choice and $I_{35}$ is another consistent choice with this maximal
cut.  We have checked that more than 700 kinematic solutions of this
cut fail to distinguish between the possibilities.  To resolve this
situation we use less degenerate kinematics with fewer cut
conditions imposed. The two cuts in \fig{nonsingletFigure}, for
example, resolve this ambiguity.

This type of ambiguity can even affect combinations of integrals
with different sets of propagators. As a nontrivial example, the
following combination of integrals vanishes in all maximal cuts we
have evaluated, other than the ones in \fig{nonsingletFigure} and
the maximal cut of $I_{50}$:
\begin{equation}
(I_{21}+I_{22}-I_{27}+I_{31}+I_{32}+I_{33}+I_{34}
 -I_{35}-I_{36}+I_{38})\Bigl|_\mathrm{cut}=0 \,.
\label{PotentialAmbiguity}
\end{equation}
This equation as well as eq.~(\ref{PotentialAmbiguity2}) should be
interpreted as a recipe for determining a combination of terms that
can vanish in a maximal cut; if a cut picks up any integral or its
permutation it should be included.  (Note that the signs shown in
\figs{cubicsFigure}{quarticsFigure} are included in the definition of
these integrals.) Note that eq.~(\ref{PotentialAmbiguity2}) is the
same ambiguity as eq.~(\ref{PotentialAmbiguity}), but restricted to
cuts of $I_{21}$'s topology. Other than (\ref{PotentialAmbiguity}) we
have found no ambiguity that holds for all singlet solutions of 
a maximal cut. In any case, it can be
resolved by using cuts with fewer on-shell conditions.  In particular,
the cuts in \fig{nonsingletFigure} resolve the ambiguity
(\ref{PotentialAmbiguity}).  These cuts are only consistent with
integrals $I_{21}$, $I_{22}$, $I_{27}$, $I_{31}$, $I_{32}$, $I_{33}$
and $I_{34}$ included in the amplitude and $I_{35}$, $I_{36}$ and
$I_{38}$ excluded.  It is likely that this kind of ambiguity also
exists at all higher loops when using maximal cuts, but again it should be
resolved by using less-restrictive cuts.

Although it is straightforward to solve analytically for any given
kinematic configuration, it can get quite tedious since many of the
pseudo-conformal integrals have well over 100 singlet cuts each.  It
is therefore simpler to do so numerically. The bispinor formalism,
which is automatically on-shell, enables us to choose which
kinematic solution to solve for numerically : Our procedure is to first
assign spinors, one to each three-point vertex, with $\lambda$'s
assigned to each \MinusVertex{} vertex and $\tlambda$'s assigned to
each \PlusVertex{} blob.  If two nearest-neighbor three-point vertices
are of the \MinusVertex{} type, the two $\lambda$ spinors of the
vertices are set equal to each other.  The on-shell constraints force the
$\lambda$ spinors to be proportional to each other, but we use the
rescaling freedom of the spinors, mentioned below \eqn{solution}, to
set the proportionality constant to unity.  Similarly, if two
nearest-neighbor three-point vertices are of the \PlusVertex{} type,
the $\tlambda$ spinors are set equal to each other.  This gives us a
list of $\lambda$'s and $\widetilde{\lambda}$'s that uniquely
specifies the solution, but their values are not yet determined.  The
momentum of a cut propagator between blobs of opposite sign is given
by taking the tensor product of the two spinors associated with the
blobs the propagator connects to.  One must also allow for a complex
scale factor multiplying each momentum between these blobs since it is not
possible to simultaneously remove the proportionality constants in
both $\lambda$ and $\tlambda$.
We solve the momentum conservation constraints by
numerically minimizing the sum of the squares of absolute value of all
momentum conservation relations that should vanish.  At the solution
point this vanishes.  

In some cases the numerical convergence is insufficiently fast.  If
necessary a given cut can always be analyzed analytically. But it is
simplest to discard unstable or poorly convergent solutions because
there are plenty of other solutions available.  We have performed on
the order of 100 singlet maximal cuts corresponding to each of the
propagator configurations of the pseudo-conformal basis integrals with
only three-point vertices, effectively exhausting the singlets. For
the integrals also containing four-point vertices, we have performed
on the order of 10 singlet maximal cuts, again effectively exhausting
the singlets.  We have also checked a handful of non-singlet
solutions, in particular for those diagrams which have no singlet
solutions. In all cases, we find that the cuts are consistent with the
ansatz (\ref{FiveLoopAnsatz}).

\section{Cross Checks on Coefficients From Two- and Three-Particle
Generalized Cuts}
\label{ConfirmingCutsSection}

The kinematics used in the maximal cuts is rather restricted, so
additional checks are desirable.  We have evaluated two such generalized cuts
in four dimensions.  In $D$ dimensions we evaluated various
two-particle cuts.  These cuts also provide a confirmation that no
other integrals appear in the amplitudes besides the pseudo-conformal
ones.

\subsection{Cuts in four dimension}

\begin{figure}
\centerline{\epsfxsize 3.8 truein \epsfbox{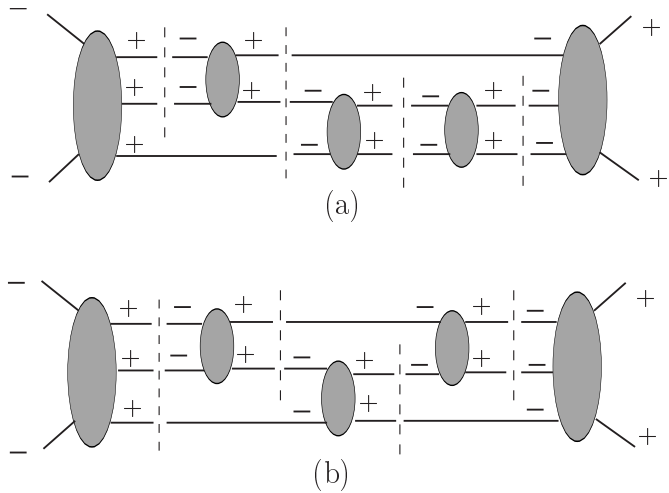}}
\caption{The two singlet cuts containing only gluons in $D=4$.}
 \label{FiveLoopCutsFigure}
\end{figure}

The easiest four-dimensional cuts to evaluate are the singlet cuts,
involving only MHV gluon tree amplitudes and a single three-particle
cut, shown in \fig{heptacutFigure}. As described in
\sect{UnitaritySection}, the non-$\gamma_5$ terms, obtained after
dividing by the tree amplitude (see \eqn{TwoLoopSingletCut}) give
precisely half the cut of the final amplitude at least through four
loops. By evaluating the cut of \fig{FiveLoopCutsFigure}, we obtain a
non-trivial check, since we find that a similar result holds at five
loops.

Our evaluation confirms agreement of the non-$\gamma_5$ terms in the
singlet cut of \fig{FiveLoopCutsFigure}(a) with 1/2 the value of the
unintegrated corresponding cut of the ansatz (\ref{FiveLoopAnsatz}).
This provides a non-trivial confirmation that we have properly
determined the coefficient of integrals\,
\def\hs{\hskip .1cm}
\begin{equation}
I_3, \hs I_8,\hs I_9, \hs I_{10},\hs I_{13},\hs I_{14},
\hs I_{17},\hs I_{18}, \hbox{ and }I_{28}\,,
\end{equation}
from the maximal cuts.
Similarly, we have confirmed that the non-$\gamma_5$ terms
in the singlet cut (b) agrees with $1/2$ times the value
of the corresponding cut of the ansatz (\ref{FiveLoopAnsatz}). This
checks that the coefficients of the integrals,
\begin{equation}
I_2, \hs I_3, \hs I_6, \hs I_{9}, \ldots ,I_{12}, \hs I_{17}, \hs I_{18},\hs
I_{25},\hbox{ and }I_{30}  \,,
\end{equation}
are also correct.  Moreover this also checks that integrals which
have cuts of the forms in \fig{FiveLoopCutsFigure}, but are not
pseudo-conformal do not appear in the amplitude.

\subsection{Cuts in $D$ dimensions}

A more rigorous check comes the evaluation of the $D$-dimensional
cuts.  As already mentioned, beyond one loop, no theorem has been
proven that four-dimension cuts are sufficient for determining
complete amplitudes in supersymmetric theories.  It is therefore
important to evaluate at least some cuts in $D$ dimensions. This is
especially true if we wish to apply the results away from $D=4$.

We evaluate the $D$-dimensional cuts of MSYM, by interpreting it
instead as ten dimensional $\NeqOne$ supersymmetric Yang-Mills,
dimensionally reduced to $D$ dimensions.  As mentioned in
\sect{UnitaritySection}, this way of evaluating the MSYM amplitudes
has the advantage of simplifying the bookkeeping on which states are
present: the $\NeqOne$ multiplet consists of only a single gluon and
gluino, each of which is composed of $8N_c$ degrees of freedom.  With
this formulation, all states are included, using $D$-dimensional
momenta in the cuts.

The simplest class of integrals to check in $D$ dimensions are ones
which can be constructed by iterating two-particle cuts, following
the discussion of refs.~\cite{BRY,BDDPR}.  The two-particle sewing
equation, which is valid in $D$ dimensions, is,
\begin{equation}
\sum_{\NeqFour \rm\ states}\hskip -.2 cm
 A_4^{\tree}(l_1, 1,2, l_2) \, A_4^\tree(-l_2, 3,4, -l_1)
 = - i\, A_4^\tree(1,2,3,4) \, {s t  \over (l_1 - k_1)^2 (l_1 + k_4)^2} \,.
\label{TwoParticleSewing}
\end{equation}
Since the tree amplitude $A_4^\tree$ appears on the right-hand-side,
the same two-particle sewing algebra appears at the next loop order.
The iterated two-particle cuts allow us to confirm
that the coefficients of integrals
\def\hs{\hskip .1cm}
\begin{equation}
I_1 \ldots I_{10}, \hs I_{15}, \hs I_{16}, \hs I_{19}, \hs I_{20},
 \hs I_{41} \ldots I_{45},\hs I_{47}, \hs I_{53}, \hs I_{57}, \hs I_{58} \,,
\label{IteratedTwoParticleList}
\end{equation}
have all been determined correctly.

\begin{figure}
\centerline{\epsfxsize 5 truein \epsfbox{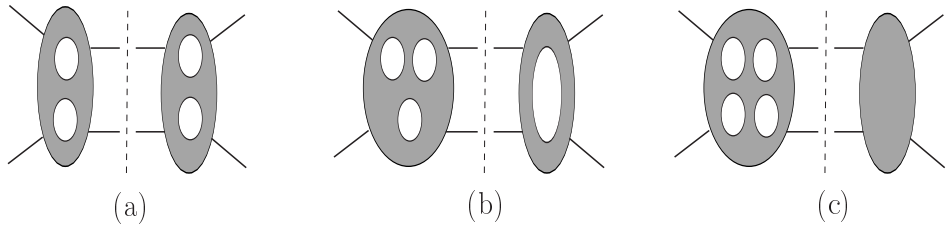}}
\caption{The $D$-dimensional two-particle cut dividing the five-loop
amplitude into 
(a) two two-loop
amplitudes, (b) a one-loop and three-loop amplitude and
(c) a four-loop and a tree amplitude.
All physical states are summed over in the cuts.}
 \label{FiveLoop2PartDCutFigure}
\end{figure}

We have also checked the $D$-dimensional two-particle cuts which split
the five-loop four-point amplitude into a product of a two two-loop
amplitudes, a one-loop and three-loop amplitude and a four-loop and a
tree amplitude, as depicted in \fig{FiveLoop2PartDCutFigure}.  Using
$D$-dimensional cuts we have evaluated the coefficients of all
integrals appearing in two- and three-loop
amplitudes, leaving the external legs in $D$ dimensions.  These
are all proportional to the $D$-dimensional tree amplitude.
We may likewise use the $D$-dimensional four-loop amplitude subject
to the same assumptions make in ref.~\cite{BCDKS}, namely the absence
of certain triangle subintegrals and the appearance of the tree-level
kinematic tensor as an overall coefficient.
We can then apply the two-particle cut
sewing equation (\ref{TwoParticleSewing}) to confirm the coefficients
of various five-loop integrals.  This allows us to provide additional
checks via $D$-dimensional unitarity that integrals,
\def\hs{\hskip .1cm}
\begin{eqnarray}
&& I_1 \ldots I_{10},  \hs I_{15} \ldots I_{20},
\hs I_{23},\hs I_{25}, \hs I_{26}, \hs I_{41} \ldots I_{47}, \hs
I_{51} \ldots \hs I_{58} \,,
\end{eqnarray}
all have the coefficients presented in \sect{MaximalCutSection}.

To have a complete proof that the ansatz (\ref{FiveLoopAnsatz}) is
complete, one would need to to confirm from $D$-dimensional unitarity
that these remaining integrals enter with the coefficients determined
in \sect{MaximalCutSection} and that there are no other
(non-conformal) integrals present.  We leave this for future work.

In general, it is rather surprising that four-dimensional unitarity
cuts are sufficient to determine the amplitudes in all dimensions.
The maximally supersymmetric theory, however, is special. Our
$D$-dimensional study here provides non-trivial evidence that at least
at four points, the four-dimensional cuts suffice.  This result may be
understood as a direct consequence of only pseudo-conformal integrals
being present, with coefficients independent of the number of
dimensions.


\section{Conclusions}
\label{ConclusionSection}

In this paper we presented an ansatz for the five-loop four-point
planar amplitude of maximally supersymmetric Yang-Mills amplitudes in
terms of a set of pseudo-conformal integrals~\cite{DHSS,BCDKS}.  We
introduced a method based on cutting the maximal number of
propagators~\cite{BCFUnitarity,BuchbinderCachazo} in each integral, to
determine very efficiently the coefficients of the integrals as they
appear in the amplitude.  We then used generalized
unitarity~\cite{GeneralizedUnitarity} with less restrictive cuts, both
in four and $D$ dimensions, to verify the correctness of the
expressions determined in this way.

Our ansatz for the planar five-loop four-point amplitude relies on a
basis of pseudo-conformal integrals, and assumes that the amplitude
can be expressed entirely in terms of such integrals.  These integrals
are the dimensionally-regulated counterparts of off-shell conformal
integrals~\cite{DHSS,BCDKS}, limited to those which have
logarithmically-divergent on-shell limits.  This assumption has been
tested and confirmed by explicit calculation through four loops.  The
assumption provides a compact basis of (plausibly independent)
integrals, and reduces the problem of computing the amplitude to that
of determining the coefficients of each integral.  We have provided
strong evidence that this continues to hold through at least five
loops, through the evaluation of a large variety of generalized
unitarity cuts, including ones evaluated in $D$ dimensions.  The
computation of additional cuts in $D$ dimensions would make it
possible to prove that our expression is indeed complete.
Alternatively, we may wonder whether it is possible to link the
conformal invariance of the theory to the absence of non-conformal
integrals.   An important cross-check would come from
showing that the infrared
singularities of the amplitude have the predicted 
form~\cite{MagneaSterman, BDS}.

The set of integrals that appears in the expression for an
$L$-loop MSYM amplitude is a subset of all pseudo-conformal
integrals.  It is interesting that the integrals which do appear, 
do so with coefficients $\pm 1$.  We presented heuristic rules
which give a partial understanding of the signs of these coefficients.
It would of course be very useful to have a complete set of heuristic
rules for predicting all signs and zeroes to arbitrary loop order.

\begin{figure}[t]
\centerline{\epsfxsize 5 truein \epsfbox{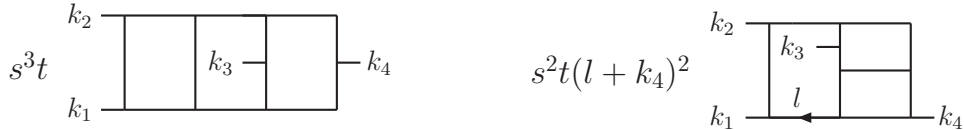}}
\caption{Non-planar examples of three-loop integrals
confirmed by cutting all the propagators. These agree with the results of
ref.~\cite{BDDPR}.}
\label{nonplanarFigure}
\end{figure}

This maximal form of generalized unitarity we employed
should also prove useful for
determining non-planar contributions.  For example, the 
non-planar contributions to the subleading-color three-loop amplitude
shown in \fig{nonplanarFigure} are easily
determined from maximal cuts. These contributions are in agreement
with known results~\cite{BDDPR,ThreeLoopNEqEight}. (Note that with the
cut conditions imposed $(l+k_4)^2$ and $2 l \cdot k_4$ are
indistinguishable in the second integral of \fig{nonplanarFigure}, so
other cuts are necessary to determine the proper factor.)

Our determination of coefficients also relied on special properties of
the four-point amplitude.  How can one compute amplitudes with a
larger number of external legs?  While some extension to the
techniques presented in the present paper will certainly be necessary,
they provide a very good starting point.  In the planar two-loop
five-point amplitude~\cite{BRYProceedings,FivePtTwoLoop}, for example,
terms with even parity relative to the tree amplitude also appear to
be expressible purely in terms of pseudo-conformal integrals. The
parity-odd terms require further study.

Beyond computations of gauge-theory amplitudes, the maximal-cut
method described here should also be useful in 
higher-loop studies of quantum gravity.  Recent
calculations have established~\cite{ThreeLoopNEqEight} that
the three-loop degree of divergence in four dimensions
(or equivalently the critical dimension)
of $\NeqEight$ supergravity is --- contrary to widely-held expectations ---
the same as that of $\NeqFour$ supersymmetric gauge theory.
There are other indications that the supergravity theory may even
be ultraviolet finite beyond three loops~\cite{GravityCancel,
GravityFinite, KITPTalk, StringFinite, ThreeLoopNEqEight}.  
These investigations point to the need for higher-loop 
computations of supergravity amplitudes, in order to
establish the critical dimension in which they first become
ultraviolet divergent.  In the approach advocated in
ref.~\cite{BDDPR}, cuts of MSYM gauge-theory amplitudes can be used
to construct cuts of $\NeqEight$ supergravity amplitudes.  The present
paper provides the required planar amplitudes at five loops.  The
non-planar contributions are more difficult, but should be within
reach.  This task would be considerably easier if a
non-planar analog of the pseudo-conformal integrals were identified.

Our expression for the planar five-loop four-point MSYM amplitude
presented in this paper has two obvious applications.  The first would
be the extraction of the planar five-loop cusp anomalous dimension,
allowing a further check of the conjectures of refs.~\cite{BCDKS,
BES}.  Another application would be a five-loop check of the iterative
structure of the amplitude.  This would provide a rather strong check
of the all-loop resummation of maximally helicity violating amplitudes
proposed in refs.~\cite{ABDK, BDS}, and help reinforce a link to a
recent string-side computation of gluon
amplitudes~\cite{AldayMaldacena}.  The latter computation, together
with all-loop-order resummations, opens a fresh venue for quantitative
studies of the AdS/CFT correspondence.

\section*{Acknowledgments}

We are grateful to Lance Dixon for many valuable discussions and
suggestions.  We also thank Radu Roiban, Emery Sokatchev, Marcus
Spradlin, and Anastasia Volovich for helpful discussions.  We thank
Academic Technology Services at UCLA for computer support.  This
research was supported in part by the US Department of Energy under
contract DE--FG03--91ER40662, and in part by the Swiss National
Science Foundation (SNF).  The figures were generated using
Jaxodraw~\cite{Jaxo}, based on Axodraw~\cite{Axo}.



\begin{thebibliography}{99}

\bibitem{Maldacena}
J.~M.~Maldacena,
Adv.\ Theor.\ Math.\ Phys.\ {\bf 2}, 231 (1998)
[Int.\ J.\ Theor.\ Phys.\ {\bf 38}, 1113 (1999)]
[hep-th/9711200];\\
%
S.~S.~Gubser, I.~R.~Klebanov and A.~M.~Polyakov,
Phys.\ Lett.\ B {\bf 428}, 105 (1998)
[hep-th/9802109];\\
%
E.~Witten,
Adv.\ Theor.\ Math.\ Phys.\  {\bf 2}, 253 (1998)
[hep-th/9802150];\\
%
O.~Aharony, S.~S.~Gubser, J.~M.~Maldacena, H.~Ooguri and Y.~Oz,
Phys.\ Rept.\  {\bf 323}, 183 (2000)
[hep-th/9905111].

\bibitem{tHooft}
G.~'t Hooft,
Nucl.\ Phys.\  B {\bf 72}, 461 (1974).

\bibitem{ABDK}
C.~Anastasiou, Z.~Bern, L.~J.~Dixon and D.~A.~Kosower,
Phys.\ Rev.\ Lett.\  {\bf 91}, 251602 (2003)
[hep-th/0309040].

\bibitem{BDS}
Z.~Bern, L.~J.~Dixon and V.~A.~Smirnov,
Phys.\ Rev.\ D {\bf 72}, 085001 (2005)
[hep-th/0505205].

\bibitem{Integrability}
J.~A.~Minahan and K.~Zarembo,
JHEP {\bf 0303}, 013 (2003)
[hep-th/0212208].
%
N.~Beisert, C.~Kristjansen and M.~Staudacher,
Nucl.\ Phys.\ B {\bf 664}, 131 (2003)
[hep-th/0303060];\\
%
A.~V.~Belitsky, A.~S.~Gorsky and G.~P.~Korchemsky,
Nucl.\ Phys.\ B {\bf 667}, 3 (2003)
[hep-th/0304028];\\
%
N.~Beisert,
Nucl.\ Phys.\ B {\bf 676}, 3 (2004)
[hep-th/0307015];\\
%
N.~Beisert,
JHEP {\bf 0309}, 062 (2003)
[hep-th/0308074];\\
%
N.~Beisert and M.~Staudacher,
Nucl.\ Phys.\ B {\bf 670}, 439 (2003)
[hep-th/0307042];\\
%
L.~Dolan, C.~R.~Nappi and E.~Witten,
JHEP {\bf 0310}, 017 (2003)
[hep-th/0308089].

\bibitem{EdenStaudacher}
B.~Eden and M.~Staudacher,
J.\ Stat.\ Mech.\  {\bf 0611}, P014 (2006)
[hep-th/0603157].

\bibitem{BES}
N.~Beisert, B.~Eden and M.~Staudacher,
J.\ Stat.\ Mech.\  {\bf 0701}, P021 (2007)
[hep-th/0610251].

\bibitem{HiddenBeauty}
F.~Cachazo, M.~Spradlin and A.~Volovich,
JHEP {\bf 0607}, 007 (2006)
[hep-th/0601031].

\bibitem{FivePtTwoLoop}
F.~Cachazo, M.~Spradlin and A.~Volovich,
Phys.\ Rev.\ D {\bf 74}, 045020 (2006)
[hep-th/0602228];\\
%
Z.~Bern, M.~Czakon, D.~A.~Kosower, R.~Roiban and V.~A.~Smirnov,
Phys.\ Rev.\ Lett.\  {\bf 97}, 181601 (2006)
[hep-th/0604074].

\bibitem{BCDKS}
Z.~Bern, M.~Czakon, L.~J.~Dixon, D.~A.~Kosower and V.~A.~Smirnov,
hep-th/0610248.

\bibitem{CSVFourLoop}
F.~Cachazo, M.~Spradlin and A.~Volovich,
hep-th/0612309.

\bibitem{AldayMaldacena}
L.~F.~Alday and J.~Maldacena,
arXiv:0705.0303 [hep-th].

\bibitem{KLOV}
A.~V.~Kotikov, L.~N.~Lipatov and V.~N.~Velizhanin,
Phys.\ Lett.\ B {\bf 557}, 114 (2003)
[hep-ph/0301021];\\
%
A.~V.~Kotikov, L.~N.~Lipatov, A.~I.~Onishchenko and V.~N.~Velizhanin,
Phys.\ Lett.\ B {\bf 595}, 521 (2004)
[hep-th/0404092].

\bibitem{KLV}
A.~V.~Kotikov, L.~N.~Lipatov and V.~N.~Velizhanin,
Phys.\ Lett.\ B {\bf 557}, 114 (2003)
[hep-ph/0301021].

\bibitem{Klebanov}
M.~K.~Benna, S.~Benvenuti, I.~R.~Klebanov and A.~Scardicchio,
Phys.\ Rev.\ Lett.\  {\bf 98}, 131603 (2007)
[hep-th/0611135];\\
%
A.~V.~Kotikov and L.~N.~Lipatov,
Nucl.\ Phys.\  B {\bf 769}, 217 (2007)
[hep-th/0611204];\\
%
J.~Maldacena and I.~Swanson,
hep-th/0612079;\\
%
L.~F.~Alday, G.~Arutyunov, M.~K.~Benna, B.~Eden and I.~R.~Klebanov,
hep-th/0702028;\\
%
I.~Kostov, D.~Serban and D.~Volin,
hep-th/0703031;\\
%
M.~Beccaria, G.~F.~De Angelis and V.~Forini,
hep-th/0703131.

\bibitem{StrongCouplingLeading}
S.~S.~Gubser, I.~R.~Klebanov and A.~M.~Polyakov,
Nucl.\ Phys.\ B {\bf 636}, 99 (2002)
[hep-th/0204051];\\
%
S.~Frolov and A.~A.~Tseytlin,
JHEP {\bf 0206}, 007 (2002)
[hep-th/0204226].

\bibitem{NeqFourOneLoop}
Z.~Bern, L.~J.~Dixon, D.~C.~Dunbar and D.~A.~Kosower,
Nucl.\ Phys.\ B {\bf 425}, 217 (1994)
[hep-ph/9403226].

\bibitem{Fusing}
Z.~Bern, L.~J.~Dixon, D.~C.~Dunbar and D.~A.~Kosower,
Nucl.\ Phys.\ B {\bf 435}, 59 (1995)
[hep-ph/9409265].

\bibitem{DDimUnitarity}
Z.~Bern and A.~G.~Morgan,
Nucl.\ Phys.\ B {\bf 467}, 479 (1996) [hep-ph/9511336];\\
%
A.~Brandhuber, S.~McNamara, B.~J.~Spence and G.~Travaglini,
JHEP {\bf 0510}, 011 (2005)
[hep-th/0506068];\\
%
C.~Anastasiou, R.~Britto, B.~Feng, Z.~Kunszt and P.~Mastrolia,
Phys.\ Lett.\  B {\bf 645}, 213 (2007)
[hep-ph/0609191];\\
%
R.~Britto and B.~Feng,
hep-ph/0612089;\\
%
C.~Anastasiou, R.~Britto, B.~Feng, Z.~Kunszt and P.~Mastrolia,
JHEP {\bf 0703}, 111 (2007)
[hep-ph/0612277].

\bibitem{OneLoopReview}
Z.~Bern, L.~J.~Dixon and D.~A.~Kosower,
Ann.\ Rev.\ Nucl.\ Part.\ Sci.\  {\bf 46}, 109 (1996)
[hep-ph/9602280].

\bibitem{DimShift}
Z.~Bern, L.~J.~Dixon, D.~C.~Dunbar and D.~A.~Kosower,
Phys.\ Lett.\ B {\bf 394}, 105 (1997)
[hep-th/9611127].

\bibitem{GeneralizedUnitarity}
Z.~Bern, L.~J.~Dixon and D.~A.~Kosower,
Nucl.\ Phys.\ B {\bf 513}, 3 (1998)
[hep-ph/9708239];\\
%
Z.~Bern, L.~J.~Dixon and D.~A.~Kosower,
JHEP {\bf 0408}, 012 (2004)
[hep-ph/0404293];\\
%
Z.~Bern, V.~Del Duca, L.~J.~Dixon and D.~A.~Kosower,
Phys.\ Rev.\ D71:045006 (2005)
[hep-th/0410224].

\bibitem{BRY}
Z.~Bern, J.~S.~Rozowsky and B.~Yan,
Phys.\ Lett.\ B {\bf 401}, 273 (1997)
[hep-ph/9702424].

\bibitem{BDDPR}
Z.~Bern, L.~J.~Dixon, D.~C.~Dunbar, M.~Perelstein and J.~S.~Rozowsky,
Nucl.\ Phys.\ B {\bf 530}, 401 (1998)
[hep-th/9802162].

\bibitem{BRYProceedings}
Z.~Bern, J.~Rozowsky and B.~Yan,
hep-ph/9706392.

\bibitem{BCFUnitarity}
R.~Britto, F.~Cachazo and B.~Feng,
Nucl.\ Phys.\  B {\bf 725}, 275 (2005)
[hep-th/0412103].

\bibitem{WittenTopologicalString}
E.~Witten,
Commun.\ Math.\ Phys.\  252:189 (2004)
[hep-th/0312171].

\bibitem{BuchbinderCachazo}
E.~I.~Buchbinder and F.~Cachazo,
JHEP {\bf 0511}, 036 (2005)
[hep-th/0506126].

\bibitem{DHSS}
J.~M.~Drummond, J.~Henn, V.~A.~Smirnov and E.~Sokatchev,
JHEP {\bf 0701}, 064 (2007)
[hep-th/0607160].

\bibitem{VolodyaIntegrals}
V.~A.~Smirnov,
Phys.\ Lett.\ B {\bf 491}, 130 (2000)
[hep-ph/0007032];\\
%
Phys.\ Lett.\ B {\bf 500}, 330 (2001)
[hep-ph/0011056];\\
%
Phys.\ Lett.\ B {\bf 524}, 129 (2002)
[hep-ph/0111160];\\
%
V.~A.~Smirnov, {\it Evaluating Feynman integrals},
Springer tracts in modern physics, {\bf 211}
(Springer, Berlin, Heidelberg, 2004).

\bibitem{MB}
M.~Czakon,
Comput.\ Phys.\ Commun.\  {\bf 175}, 559 (2006)
[hep-ph/0511200];\\
%
C.~Anastasiou and A.~Daleo,
JHEP {\bf 0610}, 031 (2006)
[hep-ph/0511176].

\bibitem{GravityCancel}
Z.~Bern, L.~J.~Dixon, M.~Perelstein and J.~S.~Rozowsky,
Nucl.\ Phys.\ B {\bf 546}, 423 (1999)
[hep-th/9811140];\\
%
Z.~Bern, N.~E.~J.~Bjerrum-Bohr and D.~C.~Dunbar,
JHEP {\bf 0505}, 056 (2005)
[hep-th/0501137];\\
%
N.~E.~J.~Bjerrum-Bohr, D.~C.~Dunbar and H.~Ita,
Phys.\ Lett.\ B {\bf 621}, 183 (2005)
[hep-th/0503102];\\
%
N.~E.~J.~Bjerrum-Bohr, D.~C.~Dunbar, H.~Ita, W.~B.~Perkins and K.~Risager,
JHEP {\bf 0612}, 072 (2006)
[hep-th/0610043].

\bibitem{GravityFinite}
Z.~Bern, L.~J.~Dixon and R.~Roiban,
Phys.\ Lett.\  B {\bf 644}, 265 (2007)
[hep-th/0611086].

\bibitem{ThreeLoopNEqEight}
Z.~Bern, J.~J.~Carrasco, L.~J.~Dixon, H.~Johansson, D.~A.~Kosower
and R.~Roiban,
hep-th/0702112.

\bibitem{KITPTalk}
Z. Bern, presented at {\it KITP Workshop: Mathematical
Structures in String Theory}, Aug. 1-Dec. 16, 2005,
http://online.kitp.ucsb.edu/online/strings05/bern;\\
Z. Bern, presented at {\it The 2006 Niels Bohr Summer Institute: Frontiers in
Theoretical Particle Physics}, http://www.nbsi.nbi.dk/bern

\bibitem{StringFinite}
G.~Chalmers,
hep-th/0008162;\\
%
M.~B.~Green, J.~G.~Russo and P.~Vanhove,
JHEP {\bf 0702}, 099 (2007)
[hep-th/0610299];\\
%
M.~B.~Green, J.~G.~Russo and P.~Vanhove,
Phys.\ Rev.\ Lett.\  {\bf 98}, 131602 (2007)
[hep-th/0611273].

\bibitem{OoguriPrivate}
M.~B. Green, H. Ooguri and J.~H. Schwarz, private communications;
arXiv:0704.0777 [hep-th].

\bibitem{KLT}
H.~Kawai, D.~C.~Lewellen and S.~H.~H.~Tye,
Nucl.\ Phys.\ B {\bf 269}, 1 (1986).

\bibitem{TreeReview}
M.~L.~Mangano and S.~J.~Parke,
Phys.\ Rept.\ {\bf 200}, 301 (1991);\\
%
L.~J.~Dixon,
in {\it QCD \& Beyond: Proceedings of TASI '95},
ed. D.~E.~Soper (World Scientific, 1996)
[hep-ph/9601359].

\bibitem{SWI}
M.~T.~Grisaru, H.~N.~Pendleton and P.~van Nieuwenhuizen,
Phys.\ Rev.\ D {\bf 15}, 996 (1977);\\
%
M.~T.~Grisaru and H.~N.~Pendleton,
Nucl.\ Phys.\ B {\bf 124}, 81 (1977);\\
%
S.~J.~Parke and T.~R.~Taylor,
Phys.\ Lett.\ B {\bf 157}, 81 (1985)
[Erratum-ibid.\  {\bf 174B}, 465 (1986)];\\
%
Z.~Kunszt,
Nucl.\ Phys.\ B {\bf 271}, 333 (1986).

\bibitem{SpinorHelicity}
F.~A.~Berends, R.~Kleiss, P.~De Causmaecker, R.~Gastmans and T.~T.~Wu,
Phys.\ Lett.\ B103:124 (1981);\\
%
P.~De Causmaecker, R.~Gastmans, W.~Troost and T.~T.~Wu,
Nucl.\ Phys.\ B206:53 (1982);\\
%
Z.~Xu, D.~H.~Zhang and L.~Chang,
TUTP-84/3-TSINGHUA;\\
%
R.~Kleiss and W.~J.~Stirling,
Nucl.\ Phys.\ B262:235 (1985);\\
%
J.~F.~Gunion and Z.~Kunszt,
Phys.\ Lett.\ B161:333 (1985);\\
%
Z.~Xu, D.~H.~Zhang and L.~Chang,
Nucl.\ Phys.\ B291:392 (1987).

\bibitem{MagneaSterman}
L.~Magnea and G.~Sterman,
Phys.\ Rev.\ D {\bf 42}, 4222 (1990).

\bibitem{FDH}
Z.~Bern and D.~A.~Kosower,
Nucl.\ Phys.\ B {\bf 379}, 451 (1992);\\
%
Z.~Bern, A.~De Freitas, L.~Dixon and H.~L.~Wong,
Phys.\ Rev.\ D {\bf 66}, 085002 (2002)
[hep-ph/0202271];\\
%
A.~De Freitas and Z.~Bern,
JHEP {\bf 0409}, 039 (2004)
[hep-ph/0409007].

\bibitem{Siegel}
W.~Siegel,
Phys.\ Lett.\ B {\bf 84}, 193 (1979).

\bibitem{Nakanishi}
N.~Nakanishi, {\it Graph Theory and Feynman Integrals} (Gordon and
Breach, New York, 1971).

\bibitem{HoweStelleNew}
P.~S.~Howe and K.~S.~Stelle,
Phys.\ Lett.\ B {\bf 554}, 190 (2003)
[hep-th/0211279].

\bibitem{BCFW}
R.~Britto, F.~Cachazo, B.~Feng and E.~Witten,
Phys.\ Rev.\ Lett.\ 94:181602 (2005)
[hep-th/0501052].

\bibitem{EarlyGeneralizedUnitarity}
R.~J.~Eden, P.~V.~Landshoff, D.~I.~Olive, J.~C.~Polkinghorne, {\it
The Analytic S Matrix} (Cambridge University Press, 1966).

\bibitem{QQGGG}
Z.~Bern, L.~J.~Dixon and D.~A.~Kosower,
Nucl.\ Phys.\  B {\bf 437}, 259 (1995)
[hep-ph/9409393].

\bibitem{RSV}
R.~Roiban, M.~Spradlin and A.~Volovich,
Phys.\ Rev.\  D {\bf 70}, 026009 (2004)
[hep-th/0403190].

\bibitem{Vergu}
C.~Vergu,
Phys.\ Rev.\  D {\bf 75}, 025028 (2007)
[hep-th/0612250].

\bibitem{TwoLoopGluons}
Z.~Bern, A.~De Freitas and L.~J.~Dixon,
JHEP {\bf 0203}, 018 (2002)
[hep-ph/0201161].

\bibitem{Jaxo}
D.~Binosi and L.~Theussl,
Comput.\ Phys.\ Commun.\ {\bf 161}, 76 (2004)
[hep-ph/0309015].

\bibitem{Axo}
J.~A.~M.~Vermaseren,
Comput.\ Phys.\ Commun.\ {\bf 83}, 45 (1994).

\end{thebibliography}
\end{document}